\documentclass[a4paper,fleqn,usenatbib]{mnras}

\usepackage[T1]{fontenc}
\usepackage{ae,aecompl}

\usepackage{graphicx,epsfig}	
\usepackage{amsmath}	
\usepackage{amssymb}	
\usepackage[english]{babel}
\usepackage[justification=centering]{caption}
\usepackage{hyperref}
\usepackage{float}

\title[Relativistic reflection fraction and photon index in AGN]{Correlation between relativistic reflection fraction and photon index in \textit{NuSTAR} sample of Seyfert~1 AGN}

\author[S. H. Ezhikode et al.]{Savithri H. Ezhikode,$^{1,3}$\thanks{E-mail: savithri@iucaa.in}
Gulab C. Dewangan,$^{1}$\thanks{E-mail: gulabd@iucaa.in}
Ranjeev Misra,$^{1}$\thanks{E-mail: rmisra@iucaa.in}
\newauthor
Ninan Sajeeth Philip$^{1,2,3}$
\\
$^{1}$ Inter-University Centre for Astronomy \& Astrophysics, Post Bag 4, Ganeshkhind, Pune, India\\
$^{2}$ Artificial Intelligence Research and Intelligent Systems. Thelliyoor, Kerala - 689644, India\\
$^{3}$ Department of Physics, St. Thomas College, Kozhencherry, Kerala 689641, India\\
}

\date{Accepted 2020 April 22. Received 2020 April 6; in original form 2019 November 27}

\pubyear{2020}

\begin{document}
\label{firstpage}
\pagerange{\pageref{firstpage}--\pageref{lastpage}}
\maketitle

\begin{abstract}
The primary X-ray emission from AGN, described by a power law, irradiates the accretion disc producing reflection features in the spectrum. The reflection features arising from the inner regions of the disc can be significantly modified by the relativistic effects near the black hole. We investigate the relationship between the relativistic reflection fraction $R_{f}$, defined as the ratio of the coronal intensity that illuminates the accretion disc to the coronal intensity observed directly, and the hard X-ray photon index $\Gamma$ of a \textit{NuSTAR} sample of Seyfert~1 galaxies. The X-ray spectra are modelled using {\sc relxill} code which helps to directly obtain the reflection fraction of a relativistically smeared reflection component. The parameter $R_{f}$ depends on the amount of Comptonised X-ray emission intercepted by the inner accretion disc. We found a positive correlation between $\Gamma$ and $R_{f}$ in our sample. Seed photons from a larger area of an accretion disc entering the corona will result in increased cooling of the coronal plasma, giving rise to steeper X-ray spectrum. The corona irradiating the larger area of the disc will result in higher reflection fraction. Thus, the observed $R_{f}-\Gamma$ relation is most likely related to the variations in the disc-corona geometry of AGN.
\end{abstract}

\begin{keywords}
galaxies:active -- galaxies:Seyfert -- X-rays:galaxies
\end{keywords}



\section{Introduction}
\label{Sec1}

The X-ray spectra of Seyfert~1 galaxies are composed of several components, apart from the underlying power law continuum. A portion of the primary X-ray power law photons, believed to be produced by the Comptonisation of the optical/UV photons from the accretion disc by the hot electrons in the corona (\citet{1993ApJ...413..507H}), irradiate the accretion disc and the circumnuclear material. This primary continuum gets reprocessed producing the reflection features in the X-ray band (e.g. \citet{1991MNRAS.249..352G,1991A&A...247...25M,1999ASPC..161..178R, 2010SSRv..157..167F}), making a complex broadband spectrum. 

The hard X-ray spectrum of AGN consists of a hump-like feature around 20$-$40~keV. This reflection hump is produced as a result of the interaction of the primary X-ray photons with the cold gas. The hard X-ray photons irradiating the optically-thick material are Compton backscattered whereas the softer X-rays are photoelectrically absorbed \citep{1990MNRAS.242..660N, 1994MNRAS.268..405N, 1988ApJ...335...57L, 2010SSRv..157..167F}, leading to the reflection hump. At energies below 20~keV, there is a deficit in the spectrum due to the absorption mainly by iron. In addition, the incident X-rays can be reprocessed into fluorescent emission lines. The Fe K$\alpha$ line emission at 6.4~keV is the strongest among these. Since the reflection features are dependent on the structure, temperature, chemical composition and ionisation state of the gas, the reflection spectrum in AGN can give direct information about the physical conditions in the reflecting medium. 

The observed Fe~K$\alpha$ emission line generally consists of a narrow and a broad component. The narrow line is attributed to the emission by fluorescence from distant (and low velocity) materials like the cold torus or the outer BLR. On the other hand, the broad Fe K$\alpha$ emission line is thought to be arising from the inner part of the accretion disc. The effects of Doppler boosting, light-bending and gravitational redshift can change the width and profile of the emission lines originating from regions near the supermassive black hole (SMBH) \citep{1989MNRAS.238..729F,2000PASP..112.1145F,2003MNRAS.340L..28F,2003PhR...377..389R,2004ApJS..153..205D,
2008A&A...483..437M,2011MNRAS.416..941S,2012ApJ...755...88R,2012MNRAS.422.1914D}. Hence the broadening of Fe~K$\alpha$ emission line provides vital information on the reflection from inner regions of the accretion disc.

\begin{table*}
\begin{center}
\tabcolsep 0.2cm
\caption[Details of the \textit{NuSTAR} sample of Seyferts]{Details of the sample of Seyferts with \textit{NuSTAR} observations. The count rate is obtained for the 3$-$79~keV band.}
\scriptsize
\label{tab_sample}
\begin{tabular}{clccccccccccccccrr}
\hline
\hline
\multicolumn{1}{c}{No.} & \multicolumn{1}{l}{Object} & \multicolumn{1}{c}{RA} & \multicolumn{1}{c}{DEC} & \multicolumn{1}{c}{\textit{z}} & \multicolumn{1}{c}{Observation} & \multicolumn{1}{c}{Obs. Date} & \multicolumn{1}{c}{Exposure} & \multicolumn{1}{c}{Count Rate}\\
\\
\multicolumn{1}{c}{} & \multicolumn{1}{l}{} & \multicolumn{1}{c}{degrees} & \multicolumn{1}{c}{degrees} & \multicolumn{1}{c}{} & \multicolumn{1}{c}{ID} & \multicolumn{1}{c}{yyyy-mm-dd} & \multicolumn{1}{c}{s} & \multicolumn{1}{c}{10$^{-2}$ count s$^{-1}$}\\
\hline
1 & 3C~382 & 278.76413 & 32.69633 & 0.058 & 60001084002 & 2013-12-18 & 82585 &  81.46$\pm$0.32\\
2 & 4C~74.26 & 310.65545 & 75.13401 & 0.104 & 60001080006 & 2014-10-30 & 90925 & 74.94$\pm$0.29 \\
3 & Ark~120 & 79.04759 & -0.14983 & 0.033 & 60001044004 & 2014-03-22 & 65458 & 74.30$\pm$0.34 \\
4 & Ark~564 & 340.66394 & 29.72536 & 0.025 & 60101031002 & 2015-05-22  & 211226 & 28.60$\pm$0.12 \\
5 & ESO~141-G~055 & 290.30890 & -58.67031 & 0.037 & 60201042002 & 2016-07-15 & 93011 & 67.16$\pm$0.27\\
6 & Fairall~9 & 20.94074 & -58.80578 & 0.047 & 60001130003 & 2014-05-09 & 93838 & 57.10$\pm$0.25 \\
7 & IC~4329A & 207.33028 & -30.30944 & 0.016 & 60001045002 & 2012-08-12 & 162399 & 205.00$\pm$0.36 \\
8 & MCG-6-30-15 & 203.97378 & -34.29554 & 0.008 & 60001047003 & 2013-01-30 & 127232 & 120.00$\pm$0.31 \\
9 & Mkn~335 & 1.58134 & 20.20291 & 0.026 & 60001041005 & 2013-06-25 & 93028 & 18.00$\pm$0.15 \\
10 & Mrk~1040 & 37.06032 & 31.31166 & 0.017 & 60101002004 & 2015-08-15 & 64252 & 69.60$\pm$0.34 \\
11 & Mrk~110 & 141.30362 & 52.28626 & 0.035 & 60201025002 & 2017-01-23 & 184563 & 99.24$\pm$0.24\\
12 & Mrk~590 & 33.63984 & -0.76669 & 0.026 & 90201043002 & 2016-12-02 & 51003 & 5.74$\pm$0.11 \\
13 & Mrk~766 & 184.61046 & 29.81287 & 0.013 & 60001048002 & 2015-01-24 & 90174 & 52.30$\pm$0.25 \\
14 & Swift~J2127.4+5654 & 321.93729 & 56.94436 & 0.014 & 60001110005 & 2012-11-06 & 74583 & 82.28$\pm$0.34\\
\hline
\hline
\end{tabular}
\end{center}
\end{table*}

The observed hard X-ray spectrum of AGN is a combination of the reflection spectrum and the primary power law. Since the reflection component emerges due to the irradiation of a fraction of the primary X-rays on the accretion disc, the strength of reflection can be obtained from the ratio of the fluxes from the reflected emission and the direct emission. A parameter for the reflection fraction for relativistic case was first introduced in the relativistic reflection model {\sc relxill} \citep{2014ApJ...782...76G, 2016A&A...590A..76D}. More details of this model are given in Section~\ref{sec3}. As the amount of reflection could be modified by the area of the reflector and the location of the X-ray emitting region, the reflection fraction can provide important information regarding the geometry of the accretion disc and the corona.

A number of studies have discussed the connection between the geometry of corona and the reflection. In a study of black hole candidate GX~339-4, \citet{1994PASJ...46..107U} observed a correlation between the intrinsic spectral slope and the amount of reflection which may be explained by a change in the relative geometry of the corona and accretion disc. Later, \citet{1999MNRAS.303L..11Z} found a strong correlation between the intrinsic X-ray spectral slope and the amount of Compton reflection in Seyferts. This indicates the role of the cold reflecting medium as a source of seed photons for thermal Comptonisation in the hot plasma. \citet{2015MNRAS.448..703W} discussed the effect of the geometry of the corona on the relativistically blurred X-ray reflection arising from the accretion discs of AGN. They showed that low reflection fractions (ratio of the reflected flux to the continuum flux) in AGN might be observed when a patchy corona covers a large portion ($\gtrsim$85\%) of the innermost regions of the disc and Compton scatter the reflected X-rays from this region. It has also been previously reported that the amount of reflection is stronger in low-luminosity Seyferts than the high-luminosity, high-redshift sources \citep{2000MNRAS.316..234R,2005MNRAS.364..195P,2008ApJ...682...81S}.

This work focuses on the relativistic reflection fraction $R_{f}$ of Seyfert 1 galaxies. This can be achieved by analysing the data from \textit{Nuclear Spectroscopic Telescope Array} (\textit{NuSTAR}; \citet{2013ApJ...770..103H}), a hard X-ray observatory with high sensitivity in the 3$-$79~keV band. We have used the \textit{NuSTAR} observations of many Seyferts for this study. The high-quality spectra from \textit{NuSTAR} helped to get better constraints on the coronal properties and other physical parameters of AGN, in the past few years (e.g. \citet{2014ApJ...788...61B,2015ApJ...800...62B,2015MNRAS.451.4375F,2017MNRAS.467.2566F,2015MNRAS.447.3029M, 2014MNRAS.440.2347M,2015MNRAS.447..160M,2016AN....337..490M,2017MNRAS.464.2565L,2017MNRAS.470.3239P, 2017MNRAS.466.4193T,2018A&A...614A..37T,2018JApA...39...15R,2018ApJ...867...67F,2018A&A...618A.167M, 2018MNRAS.479.2464G}). A number of works on individual AGN or sample of AGN have also focussed on the X-ray reflection features using \textit{NuSTAR} data (e.g. \citet{2015ApJ...799L..24Z,2017ApJ...837...21X,2018MNRAS.480.3689B,2018A&A...609A..42P}).

This paper is organised as follows. In Section~\ref{Sec2}, we present the sample selection and data reduction methods. Detailed modelling of individual AGN spectra is described in Section~\ref{sec3}. We discuss our results in Section~\ref{sec4}. The summary of the work is given in Section~\ref{sec5}. 

\section{Sample Selection \& Data Processing}
\label{Sec2}

We compiled the list of all AGN given in the \textit{NuSTAR} Master Catalog {\sc numaster} as available on July 2019 from {\sc HEASARC} archive. This initial sample consists of a total of 227 Seyfert type~1 (Sy~1) AGN observed in the {\sc science} mode. The sample includes Sy~1.0, Sy~1.2, Sy~1.5, Sy~1.8, Sy~1.9, Sy~1i, Sy~1h and some sources with unknown classification. We searched for the sub-classification of these sources using Simbad\footnote{\url{http://simbad.u-strasbg.fr/simbad/}; \cite{2000A&AS..143....9W}}, NED\footnote{\url{https://ned.ipac.caltech.edu/}}, HyperLeda\footnote{\url{http://leda.univ-lyon1.fr/}; \cite{2014A&A...570A..13M}}, and literature and found that there were 66 unobscured sources (Sy~1.0, Sy~1.2, \& narrow-line Seyfert~1 (NLS1)) in the sample. From this sample we selected sources with exposure time $\geq$50~ks and \textit{Swift}-BAT flux $\geq 10^{-11}$~erg~cm$^{-2}$s$^{-1}$ in order to get high-quality spectra. This resulted in a sample size of 23 sources which belong to the classes Sy~1.0, Sy~1.2 and NLS1.

The \textit{NuSTAR} data for the 23 unobscured sources were compiled and then reduced with the standard pipeline in the \textit{NuSTAR} Data Analysis Software ({\sc nustardas~v1.6.0}). For sources with multiple observations, we used the one with the longest exposure time. The data were processed for both Focal Plane Modules, FPMA and FPMB, using the calibration files taken from the \textit{NuSTAR} CALDB (version~20160731). The cleaned and calibrated Level-2 event files were created using the \texttt{nupipeline} task. Further, the software module \texttt{nuproducts} was used to generate the source and background spectra from the filtered event lists. The spectra were extracted from circular regions of different sizes depending on the source. However, the source and background spectra of each observation were extracted from regions of the same size.

\subsection{Final Sample}
\label{sec2.1}

The relativistic effects in the vicinity of SMBH lead to the broadening of the Fe K$\alpha$ lines in the AGN spectrum. Therefore, objects that have broad Fe emission lines are suitable for estimating the relativistic reflection fraction. The significance of Fe emission lines in the sample was verified by fitting the spectrum using {\sc xspec} version:~12.9.0. Cosmological parameters of H$_0$~=~70~km~s$^{-1}$~Mpc$^{-1}$, $\Omega_{\Lambda}$~=~0.73 and $\Omega_M$~=~0.27 are used throughout this paper, and the errors quoted correspond to 90\% confidence level. The spectral fitting procedure we followed for identifying the broad Fe line is explained below.

\begin{figure*}
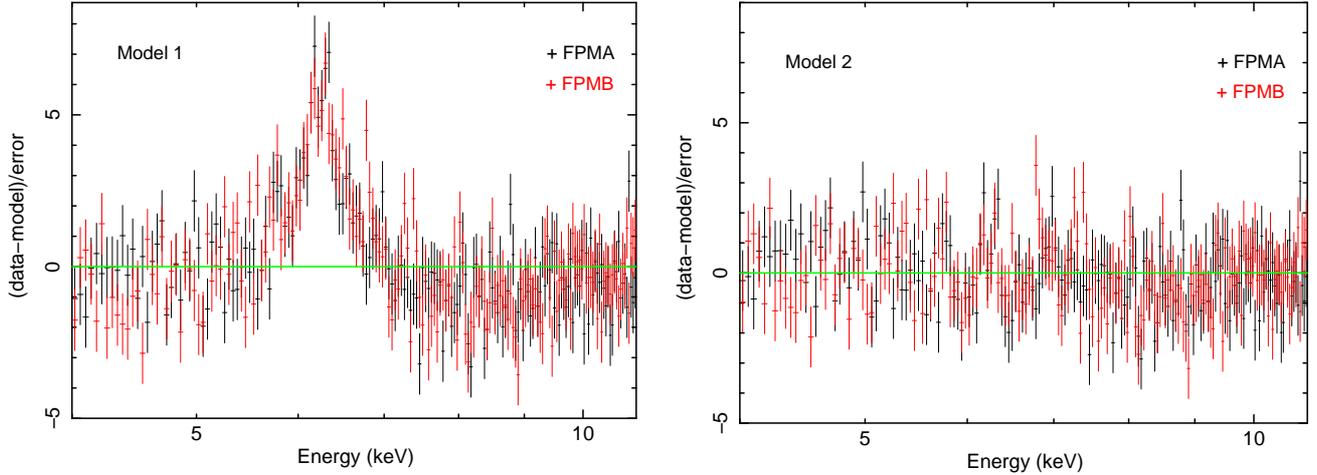

\begin{center}
\includegraphics[trim=0cm 0cm 0cm 0cm, clip=true, scale=0.32, angle=-90]{Fig1_1.ps}
\includegraphics[trim=0cm 0cm 0cm 0cm, clip=true, scale=0.32, angle=-90]{Fig1_2.ps}
\caption[Evidence for broad iron emission line in IC~4329A]{Deviations of the observed data from the best-fitting Model~1: {\sc constant$\times$tbabs$\times$powerlaw} and Model~2: {\sc constant$\times$tbabs(zgauss + powerlaw)} for IC~4329A. Left: The excess emission around $\sim 5-8$~keV showing the presence of broad iron line component. Right: The residuals from the spectral fit after the includion of a broad {\sc zgauss} component. FPMA data are plotted in black colour, and FPMB data are plotted in red colour.}
\label{Fig_residual}
\end{center}
\end{figure*}

The spectra were first fitted with the power law model ({\sc powerlaw}) corrected for Galactic absorption. We modelled the Galactic X-ray absorption with {\sc tbabs} using the Galactic column density $N_{\rm H}^{\rm Gal}$ taken from LAB Survey \citep{2005A&A...440..775K}. Since the spectra from FPMA and FPMB were fitted simultaneously, the model {\sc constant} was used for accounting the cross-calibration differences between the modules. Some sources showed  emission features around 6.4~keV. We added a \textsc{zgauss} model component in those sources, with the line energy centred at about 6.4~keV. The width of the emission line $\sigma$ was set as a free parameter. This resulted in 14 sources in which an improvement in $\chi^2$ over the power law model was observed on the addition of a broad \textsc{zgauss} ($\sigma_{\rm brd} \gtrsim 0.2$~keV). In these sources, the fit statistic improved with an \textsf{f-test} probability $<$0.01 (significant at $\sim$99\%) corresponding to a change in $\chi^2$ greater than 10. As an example, in Fig.~\ref{Fig_residual}, we show the residuals of IC~4329A before and after including the redshifted Gaussian component. The broad Fe lines were detected in the sources IC~4329A \citep{2007MNRAS.382..194N,1997ApJ...477..602N}, Mrk~766 \citep{2007MNRAS.382..194N,1997ApJ...477..602N}, MCG-6-30-15 \citep{1995Natur.375..659T,2002MNRAS.335L...1F,2001MNRAS.328L..27W,2007PASJ...59S.315M,2007MNRAS.382..194N}, Mrk~1040 \citep{1995MNRAS.276.1311R}, Mrk~335 \citep{2011MNRAS.411.2353P}, Ark~120 \citep{2007MNRAS.382..194N,2011MNRAS.410.1251N} and Fairall~9 \citep{1997ApJ...477..602N} based on the data from previous observatories.

Further, an additional \textsc{zguass} was included to check the presence of the narrow ($\sigma_{\rm nrw}=0.001$~keV) Fe K$_\alpha$ emission line ($E_{\rm nrw}$=6.4~keV) due to neutral reflection. The detection of narrow Fe emission lines by \textit{Chandra} HETG observations was already reported for the sources Fairall~9, MCG-06-30-15 (\citet{2004ApJ...604...63Y,2002ApJ...570L..47L}), IC~4329A \citep{2004ApJ...608..157M}, Ark~120 \citep{2016ApJ...832...45N}, Mrk~590 \citep{2007A&A...470...73L} and 3C~382 \citep{2007ApJ...664...88G}. In the present study, it is observed that the spectral data for three among 14 sources provided a better fit statistic on the addition of the new component. So in the final sample, there are 14 sources for which the spectra show the effect of relativistic reflection, and there are three sources in which both relativistic and distant reflection lines are clearly present in the spectra. The details of these sources are given in Table~\ref{tab_sample}, and the best-fit parameters of the three models are given in Table~\ref{tab_pow}, Table~\ref{tab_brdgau} \& Table~\ref{tab_brdnrwgau}. See Appendix~\ref{A} for the residual plots for all the sources in the sample fitted with the model {\sc tbabs$\times$powerlaw}. The sources have a redshift range of $0.008<z<0.104$. It is also noted that five sources in the final sample are NLS1 galaxies. 

\begin{table*}
\begin{center}
\scriptsize
\tabcolsep 0.4cm
\caption{Best-fit parameters of sample fitted with \textit{NuSTAR} data using the absorbed power law model. The power law normalisation $N_{\rm pow}$ is in units of photons~keV$^{-1}$cm$^{-2}$s$^{-1}$.}
\label{tab_pow}
\begin{tabular}{clccccccccccccccccrr}
\hline
\hline
\multicolumn{1}{c}{No.} & \multicolumn{1}{l}{Object} & \multicolumn{1}{c}{$N_{\rm H}^{\rm Gal}$} & \multicolumn{1}{c}{$\Gamma$} & \multicolumn{1}{c}{$N_{\rm pow}$} & \multicolumn{1}{c}{$\chi^2$/dof}\\
\\
\multicolumn{1}{c}{} & \multicolumn{1}{l}{} & \multicolumn{1}{c}{10$^{20}$cm$^{-2}$} & \multicolumn{1}{c}{} & \multicolumn{1}{c}{10$^{-2}$} & \multicolumn{1}{c}{}
\\
\hline
1 & 3C~382 & 6.98 & 1.732$\pm0.008$ & 0.76$\pm0.01$ & 762.49/672 \\
2 & 4C~74.26 & 11.60 & 1.763$\pm0.008$ & 0.78$\pm0.01$ & 925.37/669 \\
3 & Ark~120 & 9.78 & 1.863$\pm0.012$ & 12.92$^{+0.38}_{-0.37}$ & 918.55/552\\
4 & Ark~564 & 5.34 & 2.595$\pm0.010$ & 18.00$\pm0.35$ & 1098.93/657\\
5 & ESO~141-G~055 & 4.83 & 1.795$\pm0.009$ & 0.73$^{+0.02}_{-0.01}$ & 885.54/640 \\
6 & Fairall~9 & 3.16 & 1.826$\pm0.010$ & 6.61$\pm0.14$ & 878.14/597\\
7 & IC~4329A & 4.61 & 1.714$\pm0.004$ & 26.68$\pm0.27$ & 2162.95/1342\\
8 & MCG-6-30-15 & 3.92 & 1.932$\pm0.006$ & 1.71$\pm0.02$ & 1899.32/884\\
9 & Mkn~335 & 3.56 & 1.679$^{+0.020}_{-0.019}$ & 0.15$\pm0.01$ & 711.21/273\\
10 & Mrk~1040 & 6.63 & 1.751$\pm0.011$ & 7.06$\pm0.17$ & 801.90/583\\
11 & Mrk~110 & 1.30 & 1.812$\pm0.007$ & 1.16$\pm0.02$ & 598.37/493\\
12 & Mrk~590 & 2.65 & 1.655$^{+0.044}_{-0.043}$ & 0.72$\pm0.07$ & 224.25/203\\
13 & Mrk~766 & 1.78 & 2.011$\pm0.011$ & 8.89$\pm0.21$ & 799.75/532\\
14 & Swift~J2127.4+5654 & 76.50 & 1.827$\pm0.009$ & 0.98$\pm0.02$ & 1091.96/638\\
\hline
\hline
\end{tabular}
\end{center}
\vspace*{-0.2cm}
{\scriptsize Note: dof - degrees of freedom}
\end{table*}

\begin{table*}
\tabcolsep 0.2cm
\scriptsize
\begin{center}
\caption[Best-fit parameters of broad Gaussian and power law model]{Best-fit parameters of broad Gaussian and power law models. $E_{\rm brd}$, $\sigma_{\rm brd}$ and $N_{\rm brd}$ are the line energy, width and normalisation of the broad Gaussian component. The units for $N_{\rm pow}$ and $N_{\rm brd}$ are photons~keV$^{-1}$cm$^{-2}$s$^{-1}$ and  photons~cm$^{-2}$s$^{-1}$, respectively.}
\label{tab_brdgau}
\begin{tabular}{clcccccc}
\hline
\hline
\multicolumn{1}{c}{No.}	& \multicolumn{1}{l}{Object}	& \multicolumn{1}{c}{$\Gamma$}	& \multicolumn{1}{c}{N$_{\rm pow}$} & \multicolumn{1}{c}{$E_{\rm brd}$}	& \multicolumn{1}{c}{$\sigma_{\rm brd}$}	& \multicolumn{1}{c}{N$_{\rm brd}$}	& \multicolumn{1}{c}{$\chi^2$/dof}\\
\\
\multicolumn{1}{c}{}		& \multicolumn{1}{c}{}		& \multicolumn{1}{c}{}			& \multicolumn{1}{c}{10$^{-2}$}		& \multicolumn{1}{c}{keV}						& \multicolumn{1}{c}{keV}				& \multicolumn{1}{c}{10$^{-5}$}		& \multicolumn{1}{c}{}
\\
\hline
1 & 3C~382 & 1.721$\pm0.009$ & 0.74$\pm0.02$ & 6.35$\pm0.11$ & 0.43$^{+0.16}_{-0.13}$ & 4.28$^{+1.10}_{-0.95}$ & 673.49/669 \\
2 & 4C~74.26 & 1.748$\pm0.009$ & 0.74$\pm0.02$ & 6.26$\pm0.11$ & 0.44$^{+0.11}_{-0.10}$ & 5.17$^{+1.10}_{-1.00}$ & 815.20/666 \\
3 & Ark~120 & 1.849$\pm0.009$ & 1.23$\pm0.02$ & 6.36$\pm0.04$ & 0.23$\pm0.06$ & 7.00$^{+0.97}_{-0.93}$ & 807.09/649 \\
4 & Ark~564 & 2.585$^{+0.013}_{-0.015}$ & 1.66$^{+0.05}_{-0.06}$ & 5.87$^{+0.22}_{-0.30}$ & 1.30$^{+0.22}_{-0.18}$ & 9.51$^{+2.61}_{-1.76}$ & 688.08/654 \\
5 & ESO~141-G~055 & 1.786$\pm0.010$ & 0.71$^{+0.02}_{-0.01}$ & 6.34$^{+0.08}_{-0.07}$ & 0.26$^{+0.10}_{-0.12}$ & 3.08$\pm0.69$ & 788.76/637 \\
6 & Fairall~9 & 1.811$\pm0.010$ & 0.63$\pm0.01$ & 6.33$^{+0.04}_{-0.05}$ & 0.25$^{+0.06}_{-0.05}$ & 4.58$^{+0.62}_{-0.59}$ & 640.62/594 \\
7 & IC~4329A & 1.685$\pm0.003$ & 2.46$\pm0.02$ & 6.37$\pm0.03$ & 0.27$\pm0.04$ & 12.81$^{+1.08}_{-1.05}$ & 1708.03/1339 \\
8 & MCG-6-30-15 & 1.913$^{+0.008}_{-0.009}$ & 1.61$^{+0.03}_{-0.04}$ & 5.94$^{+0.16}_{-0.21}$ & 0.65$^{+0.17}_{-0.14}$ & 11.32$^{+3.03}_{-2.13}$ & 1523.16/881 \\
9 & Mkn~335 & 1.424$^{+0.037}_{-0.038}$ & 0.08$\pm0.01$ & 4.35$^{+0.24}_{-0.31}$ & 1.42$^{+0.19}_{-0.15}$ & 16.62$^{+4.10}_{-3.07}$ & 428.21/270 \\
10 & Mrk~1040 & 1.739$\pm0.012$ & 0.68$\pm0.02$ & 6.38$^{+0.08}_{-0.10}$ & 0.36$^{+0.11}_{-0.09}$ & 4.93$^{+1.01}_{-0.89}$ & 669.02/580 \\
11 & Mrk~110 & 1.807$^{+0.007}_{-0.008}$ & 1.13$\pm0.02$ & 6.37$\pm0.10$ & 0.33$^{+0.14}_{-0.11}$ & 3.37$^{+0.92}_{-0.81}$ & 531.55/490 \\
12 & Mrk~590 & 1.624$\pm0.047$ & 0.07$\pm0.01$ & 6.40$\pm0.16$ & 0.39$^{+0.16}_{-0.13}$ & 1.38$^{+0.48}_{-0.42}$ & 188.98/200 \\
13 & Mrk~766 & 2.005$\pm0.012$ & 0.87$\pm0.02$ & 6.37$^{+0.16}_{-0.17}$ & 0.43$\pm0.13$ & 2.49$^{+0.71}_{-0.68}$ & 745.02/529 \\
14 & Swift~J2127.4+5654 & 1.820$\pm0.009$ & 0.96$\pm0.02$ & 6.37$\pm0.07$ & 0.24$^{+0.09}_{-0.08}$ & 3.71$^{+0.79}_{-0.73}$ & 988.84/635 \\
\hline
\hline
\end{tabular}
\end{center}
\end{table*}

\begin{table*}
\tabcolsep 0.15cm
\scriptsize
\begin{center}
\caption[Best-fit parameters for an additional narrow Gaussian component]{Best-fit parameters of the model with an additional narrow Gaussian component. $E_{\rm brd}$ \& $N_{\rm brd}$ are the line energy and normalisation of the broad Gaussian, and $N_{\rm nrw}$ is the normalisation of the narow Gaussian component. $E_{\rm nrw}$ \& $\sigma_{\rm nrw}$ are frozen to 6.4~keV and 10$^{-3}$~keV respectively. $N_{\rm pow}$ is in units of photons~keV$^{-1}$cm$^{-2}$s$^{-1}$ and $N_{\rm brd}$ \& $N_{\rm nrw}$ are in units of photons~cm$^{-2}$s$^{-1}$.}
\label{tab_brdnrwgau}
\begin{tabular}{clccccccc}
\hline
\hline
\multicolumn{1}{c}{No.}	& \multicolumn{1}{l}{Object}	& \multicolumn{1}{c}{$\Gamma$}	&	\multicolumn{1}{c}{N$_{\rm pow}$} & \multicolumn{1}{c}{$E_{\rm brd}$}	& \multicolumn{1}{c}{$\sigma_{\rm brd}$}	& \multicolumn{1}{c}{N$_{\rm brd}$}	&  \multicolumn{1}{c}{N$_{\rm nrw}$}	& \multicolumn{1}{c}{$\chi^2$/dof}\\
\\
\multicolumn{1}{c}{}		& \multicolumn{1}{l}{}		& \multicolumn{1}{c}{}			& \multicolumn{1}{c}{10$^{-2}$}		& \multicolumn{1}{c}{keV}						& \multicolumn{1}{c}{keV}				& \multicolumn{1}{c}{10$^{-5}$}				&  \multicolumn{1}{c}{10$^{-5}$}					& \multicolumn{1}{c}{} \\ 
\hline
1 & 3C~382 & 1.721$\pm0.009$ & 0.74$\pm0.02$ & 6.34$^{+0.17}_{-0.19}$ & 0.53$^{+0.32}_{-0.19}$ & 3.88$^{+1.30}_{-1.26}$ &  $<1.39$ & 672.40/668 \\
2 & 4C~74.26 & 1.748$\pm0.009$ & 0.74$\pm0.02$ & 6.24$^{+0.12}_{-0.16}$ & 0.46$^{+0.15}_{-0.10}$ & 4.93$^{+1.18}_{-1.33}$ &  $<1.17$ & 814.97/665 \\
3 & Ark~120 & 1.847$^{+0.009}_{-0.010}$ & 1.23$\pm0.03$ & 6.27$^{+0.12}_{-0.41}$ & 0.37$^{+0.30}_{-0.14}$ & 4.95$^{+1.89}_{-1.63}$ &  2.54$^{+1.64}_{-2.06}$ & 803.34/648 \\
4 & Ark~564 & 2.583$^{+0.014}_{-0.017}$ & 1.65$^{+0.06}_{-0.07}$ & 5.75$^{+0.28}_{-0.43}$ & 1.40$^{+0.29}_{-0.22}$ & 10.02$^{+3.46}_{-2.07}$ &  0.25$\pm0.25$ & 685.31/653 \\
5 & ESO~141-G~055 & 1.786$\pm0.010$ & 0.71$^{+0.02}_{-0.01}$ & 6.34$^{+0.10}_{-0.09}$ & 0.25$^{+0.17}_{-0.12}$ & 3.09$^{+0.68}_{-1.22}$ &  $<1.08$ & 788.76/636 \\
6 & Fairall~9 & 1.779$\pm0.014$ & 0.58$\pm0.02$ & 5.00$^{+0.55}_{-1.10}$ & 1.21$^{+0.15}_{-0.16}$ & 8.57$^{+2.50}_{-5.23}$ &  2.62$\pm0.40$ & 639.40/593 \\
7 & IC~4329A$^*$ & 1.684$\pm0.003$ & 2.45$\pm0.02$ & 6.32$^{+0.07}_{-0.08}$ & 0.45$^{+0.10}_{-0.09}$ & 9.80$\pm1.57$ &  4.31$^{+1.13}_{-1.27}$ & 1687.22/1338 \\
8 & MCG-6-30-15$^*$ & 1.885$\pm0.008$ & 1.49$\pm0.03$ & $<5.03$ & 1.14$\pm0.08$ & 20.77$^{+2.82}_{-2.65}$ &  2.99$\pm0.45$ & 1394.95/880 \\
9 & Mkn~335$^*$ & 1.517$^{+0.029}_{-0.028}$ & 0.10$\pm0.01$ & $<5.02$ & 0.92$^{+0.08}_{-0.11}$ & 7.69$^{+1.08}_{-1.20}$ &  0.88$^{+0.24}_{-0.22}$ & 408.01/269 \\
10 & Mrk~1040 & 1.704$^{+0.016}_{-0.015}$ & 0.62$^{+0.03}_{-0.01}$ & $<5.24$ & 1.35$^{+0.18}_{-0.19}$ & 11.63$^{+3.62}_{-3.34}$ &  2.10$^{+0.47}_{-0.49}$ & 663.36/579 \\
11 & Mrk~110 & 1.807$\pm0.007$ & 1.13$\pm0.02$ & 6.37$^{+0.12}_{-0.14}$ & 0.33$^{+0.25}_{-0.10}$ & 3.37$^{+0.86}_{-1.28}$ &  $<1.05$ & 531.55/489 \\
12 & Mrk~590 & 1.624$^{+0.047}_{-0.046}$ & 0.07$\pm0.01$ & 6.40$^{+0.24}_{-0.18}$ & 0.40$^{+0.21}_{-0.13}$ & 1.36$^{+0.48}_{-0.69}$ &  $<0.50$ & 188.96/199 \\
13 & Mrk~766 & 2.005$\pm0.012$ & 0.87$\pm0.02$ & 6.37$^{+0.16}_{-0.19}$ & 0.42$^{+0.15}_{-0.10}$ & 2.48$^{+0.72}_{-0.80}$ &  $<0.46$ & 745.00/528 \\
14 & Swift~J2127.4+5654 & 1.819$^{+0.010}_{-0.009}$ & 0.96$\pm0.02$ & 6.34$^{+0.11}_{-0.41}$ & 0.31$^{+0.26}_{-0.16}$ & 3.05$^{+0.75}_{-1.44}$ &  $<2.12$ & 988.43/634 \\
\hline
\hline
\end{tabular}
\end{center}
\begin{scriptsize}
Note: $^*$Sources showing $|{\Delta}{\chi^2}|>10$ on the addition of a narrow Gaussian.\\
\end{scriptsize}
\end{table*}

We have also tested the presence of intrinsic absorption in the sources using the model {\sc ztbabs} parameterised by the equivalent hydrogen column $N_{\rm H}^{\rm Int}$. The fit was improved ($|\Delta \chi^2| > 10$) by the inclusion of the component for 3C~382, IC~4329A and Mrk~110. For other sources in the sample, the statistic has not improved, and the parameter $N_{\rm H}^{\rm Int}$ was not constrained. Hence the intrinsic absorption is included in further modelling for these three sources.

\section{Spectral Modelling}
\label{sec3}

Here, we use the model {\sc relxill} \citep{2014ApJ...782...76G,2014MNRAS.444L.100D} that provides a proper treatment of the relativistic reflection near strong gravitational fields of SMBH. The relativistic reflection code {\sc relxill} takes care of the angle dependent reflection spectrum, and is a combination of the models {\sc relconv} and {\sc xillver}. {\sc relconv} is the convolution model for {\sc relline}\footnote{See the website{\\}\url{http://www.sternwarte.uni-erlangen.de/~dauser/research/relline/} for more details of {\sc relline}.} code \citep{2010MNRAS.409.1534D} that is used to calculate the relativistic smearing of reflected radiation. The angle-dependent reflection code {\sc xillver} (\citet{2010ApJ...718..695G,2011ApJ...731..131G}) gives the X-ray reflected spectrum emerging from the surface of an illuminated accretion disc. The code calculates proper reflected spectrum for each point in the disc by solving the equations of radiative transfer, energy balance, and ionisation equilibrium in a Compton-thick, plane-parallel medium. The ionisation balance at each point is calculated using the photoionisation code {\sc xstar} (\citet{2001ApJS..133..221K}). 

In order to model the broad Fe line and the other spectral components, we used the local model {\sc relxill} (version~0.4c) considering the relativistic reflection in coronal geometry. The model includes both the primary continuum irradiating the accretion disc and the reflection combined with relativistic smearing. The basic parameters of this model are the indices ($\beta$1 \& $\beta$2) of the power law disc emissivity profile, break radius ($R_{\rm br}$) where emissivity profile changes from $\beta$1 to $\beta$2, black hole spin $a$, inclination with respect to the normal to the disc \textit{i}, inner and outer radii of disc $R_{\rm in}$ \& $R_{\rm out}$, ionisation parameter $\xi$ of the accretion disc, iron abundance $A_{\rm Fe}$ (in units of solar abundance) of material in the disc, photon index $\Gamma$ \& high energy cut-off $E_{\rm cut}$ of the incident power law spectrum, redshift \textit{z} of the source, reflection fraction $R_{f}$, and normalisation $N_{\rm relxill}$\footnote{The normalisations of {\sc relxill} and {\sc xillver} models are defined in Appendix~A of \cite{2016A&A...590A..76D}.}. The implementation of relativistic reflection fraction $R_{f}$, a quantity independent of inclination and the condition of the reflector, is a major advantage of {\sc relxill}. The relativistic reflection fraction is defined as the ratio of the coronal intensity that irradiates the disc to the coronal intensity that directly reaches the observer (\citet{2016A&A...590A..76D}). Unlike other reflection models (e.g. {\sc pexrav}), {\sc relxill} considers the light-bending effects near the black hole.

In the three sources where significant narrow Fe emission lines were detected (see Sec.~\ref{sec2.1}), we incoroporated a {\sc xillver} component in addition to {\sc relxill}. For this additional {\sc xillver} component, the value of log~$\xi$ was fixed at 0. This will ensure the effect of reflection from distant material resulting in neutral narrow Fe K$\alpha$ emission. The other parameters characterising the model are $\Gamma$, $A_{\rm Fe}$, $E_{\rm cut}$, \textit{i}, \textit{z} and normalisation $N_{\rm xillver}$. Thus, to model the relativistic reflection spectrum, {\sc relxill} was used and {\sc xillver} was included to model the distant reflection.

To simplify the spectral fitting, many parameters were frozen. Assuming that there is no break radius in the emissivity profile, the emissivity indices ($\beta$1 and $\beta$2) were always tied together, where $\beta$1 was free to vary. We fixed the break and outer radii to the model default values of 15~$R_{\rm g}$ and 400~$R_{\rm g}$, respectively. The inclination was set to 30$^\circ$ which is adequate for type~1 Seyferts. The spectra were fitted for different values of black hole spins, assuming that $R_{\rm in}$ extends down to the innermost stable circular orbit $R_{\rm ISCO}$. Whenever the spectrum was fit with both {\sc relxill} and {\sc xillver}, the parameters $\Gamma$, $A_{\rm Fe}$, $E_{\rm cut}$ and \textit{i} in the model components were tied.

\subsection{Fe K absorption}

\begin{figure}
\begin{center}
\includegraphics[trim=0.8cm 1.cm 0cm 0.6cm, clip=true, scale=0.26, angle=-90]{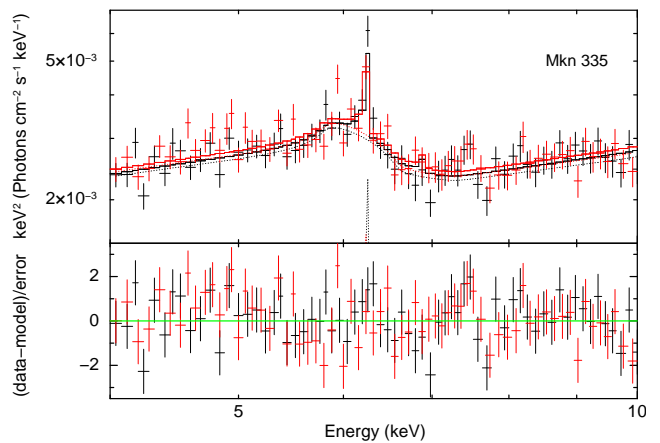}
\caption{Broadband X-ray spectral fit of Mkn~335 using the models {\sc relxill} and {\sc xillver} with spin 0 (shown in the 4--10~keV band). The upper panel shows the unfolded spectra and the best-fit model and the lower panel shows the residuals, showing narrow absorption features above $\sim$6~keV.}
\label{Fig_Mkn335_res}
\end{center}
\end{figure}

The presence of blue-shifted narrow absorption lines associated with the FeXXV-XXVI K-shell transitions have been detected in the $\sim7-10$~keV band of AGN. These features are identified with the absorption from highly ionised circumnuclear gas outflowing with velocities up to 0.2$-$0.4c. A detailed search for the evidence of such ultra-fast outflows (UFOs) in the $7-10$~keV band in a sample of radio-quiet AGN was carried out by \cite{2010A&A...521A..57T}. They detected blue-shifted Fe~K absorption lines in \textit{XMM-Newton} spectra of the sources, and modelled these lines with inverted Gaussian components by fixing the width at 10~eV or 100~eV (see Table~A2 in \cite{2010A&A...521A..57T}). They found that the corresponding outflow velocities of these blue-shifted lines range from zero to $\sim$0.3c. Their sample included nine sources in our sample. They observed single or multiple Fe K-shell absorption lines in IC~4329A ($E_{\rm rest}\sim7.69$~keV, width$=100$~eV), Ark~120 ($E_{\rm rest}\sim9.18$~keV, width$=10$~eV), Mrk~766 ($E_{\rm rest}\sim7.28$~keV, width$=100$~eV; $E_{\rm rest}\sim7.63$~keV, width$=10$~eV) and no significant absorption was detected in MCG-6-30-15, Mrk~110, Fairall~9, Mkn~335, Mrk~590 and Ark~564. In a further study, \cite{2011ApJ...742...44T} presented detailed modelling of these highly ionised absorbers using {\sc xstar} and obtained outflow velocities of $\sim0.097$c for IC~4329A, $\sim0.306$c for Ark~120 and $\sim0.082$c(0.088c) for Mrk~766. \cite{2006ApJ...646..783M} also found a narrow absorption feature in the \textit{XMM-Newton} spectrum of IC~4329A, around 7.7~keV, which is an evidence for a high velocity ($\sim0.1$c) outflow in the source. \cite{2013MNRAS.430...60G} performed a search for the Fe~K absorption lines in the \textit{Suzaku} spectra of AGN. They reported highly ionised outflow features with velocities of 0.185c, 0.231c, 0.007c and 0.061c, respectively in 4C~+74.26, SWIFT~J2127.4+5654, MCG-6-30-15 and Mrk~766. In Ark~564, \cite{2007A&A...461..931P} detected highly ionised absorption line at $\sim$8.14~keV in the \textit{XMM-Newton} EPIC-PN  spectrum, probably corresponding to the FeXXVI-K$\alpha$, indicating the presence of outflowing material with a velocity of $\sim$0.17c. However, \cite{2019MNRAS.482.5316M} reported no absorption line in Ark~564, indicating a face-on geometry of the wind. Mkn~335 has been identified with ionised absorption features in the Fe K band in \textit{XMM-Newton} and \textit{NuSTAR} spectra \citep{2019MNRAS.484.4287G, 2019ApJ...875..150L, 2014MNRAS.443.1723P}. \cite{2019MNRAS.484.4287G} observed highly ionised absorption lines in Mkn~335 which could be attributed to accretion disc winds outflowing with a velocity of $\sim0.12$c.

Although the Fe~K absorption features are reported to be present in the \textit{XMM-Newton} and/or \textit{Suzaku} spectra of most of the sources in the present sample, these narrow lines may not be detected in the low-resolution \textit{NuSTAR} spectra. However, we examined the spectra of the sources in the $\sim6-10$~keV band to identify the presence of Fe~K absorption line. The spectrum of Mkn~335 showed narrow absorption lines above $\sim6$~keV which could be related to the ultra-fast outflows (see Fig.~\ref{Fig_Mkn335_res}). We then included the analytic model {\sc warmabs} to model the outflow features, instead of using {\sc xstar mtables}. The model gives the absorber column density, ionisation parameter, element abundances in the solar unit, turbulent velocity and redshift of the outflowing gas. Despite the absorption profiles were found to be modelled, the addition of {\sc warmabs} did not improve the fit significantly over  {\sc relxill+xillver} model. Moreover, the observed redshift of the absorber corresponds to an outflow velocity of $\sim{\rm 4000~km~s}^{-1}$, which is less than the expected velocity for an ultra-fast outflow. It should also be noted that the best-fit reflection fraction and photon index for this model lie within the range obtained earlier, with lower mean values.

\begin{figure}
\begin{center}
\includegraphics[trim=0.8cm 1.2cm 0cm 0.05cm, clip=true, scale=0.26, angle=-90]{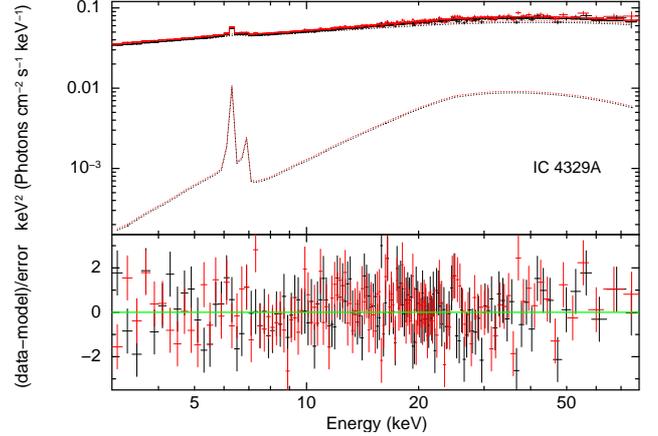}
\caption{Broadband X-ray ($3-79$~keV) spectra of IC~4329A fitted with the models {\sc relxill} and {\sc xillver} for spin 0. The upper panel shows the unfolded spectra and the best-fit models while the lower panel shows the residuals for the spectral fit.}
\label{Fig_fit}
\end{center}
\end{figure}

\begin{table*}
\scriptsize
\tabcolsep 0.1cm
\caption{Best-fit parameters of the models {\sc constant$\times$tbabs$\times$relxill} and/or {\sc constant$\times$tbabs(relxill+xillver)} for inclination \textit{i}=30$^\circ$ \& spin $a=0$. $\Gamma$: hard X-ray photon index, $\beta$1: emissivity index, log~$\xi$: logarithm of ionisation parameter, $A_{\rm Fe}$: iron abundance in solar units, $E_{\rm cut}$: cut-off energy in keV, $R_f$: reflection fraction, $N_{\rm relxill}$: Normalisation of {\sc relxill}, $N_{\rm xillver}$: Normalisation of {\sc xillver}.}
\begin{center}
\label{tab_a0}
\begin{tabular}{clcccccccccccccccccc}
\hline
\hline
\multicolumn{1}{c}{No.} & \multicolumn{1}{l}{Object} &  \multicolumn{1}{c}{$N_{\rm H}^{\rm Int}$} & \multicolumn{1}{c}{$\Gamma$} & \multicolumn{1}{c}{$E_{\rm cut}$} & \multicolumn{1}{c}{$R_f$} & \multicolumn{1}{c}{$\beta$1} & \multicolumn{1}{c}{log~$\xi$} & \multicolumn{1}{c}{$A_{\rm Fe}$} & \multicolumn{1}{c}{$N_{\rm relxill}$} & \multicolumn{1}{c}{$N_{\rm xillver}$} & \multicolumn{1}{c}{$\chi^2$/dof}\\
\\
\multicolumn{1}{c}{} & \multicolumn{1}{l}{} &  \multicolumn{1}{c}{10$^{20}{\rm cm}^{-2}$} & \multicolumn{1}{c}{} & \multicolumn{1}{c}{keV} & \multicolumn{1}{c}{} & \multicolumn{1}{c}{} & \multicolumn{1}{c}{} & \multicolumn{1}{c}{} & \multicolumn{1}{c}{10$^{-4}$} & \multicolumn{1}{c}{10$^{-5}$} & \multicolumn{1}{c}{}
\\
\hline
1 & 3C~382 & 36.34$^{+35.55}_{-22.81}$ & 1.67$\pm0.05$ & 132.75$^{+98.32}_{-39.98}$ & 0.13$^{+0.07}_{-0.03}$ & $<2.19$ & $<2.13$ & $>4.37$ & 1.71$^{+0.27}_{-0.17}$ &  & 650.07/666 \\
2 & 4C~74.26 &  & 1.81$^{+0.05}_{-0.06}$ & 155.64$^{+49.05}_{-48.21}$ & 0.58$\pm0.16$ & 2.46$^{+0.57}_{-0.43}$ & 2.70$^{+0.06}_{-0.35}$ & 0.93$^{+1.30}_{-0.27}$ & 1.34$^{+0.08}_{-0.09}$ &  & 696.34/664 \\
3 & Ark~120 &  & 1.97$\pm0.03$ & 393.17$^{+497.82}_{-152.75}$ & 0.61$^{+0.04}_{-0.10}$ & $<1.52$ & $<1.07$ & 2.43$^{+0.60}_{-0.57}$ & 2.36$^{+0.21}_{-0.15}$ &  & 693.41/647 \\
4 & Ark~564 &  & 2.31$^{+0.04}_{-0.06}$ & 59.00$^{+14.82}_{-11.84}$ & 0.59$^{+0.30}_{-0.29}$ & $<1.96$ & $>4.27$ & 3.57$^{+3.33}_{-1.75}$ & 0.73$^{+0.31}_{-0.23}$ &  & 644.43/652 \\
5 & ESO~141-G~055 &  & 1.92$^{+0.03}_{-0.02}$ & $>450.58$ & 0.53$^{+0.04}_{-0.11}$ & $<1.31$ & $<2.02$ & 1.58$^{+0.90}_{-0.44}$ & 1.72$^{+0.06}_{-0.16}$ &  & 669.75/635 \\
6 & Fairall~9 &  & 1.92$\pm0.04$ & 349.25$^{+381.40}_{-139.91}$ & 0.65$^{+0.11}_{-0.17}$ & 1.37$^{+0.57}_{-1.33}$ & $<1.09$ & 2.97$^{+0.75}_{-0.64}$ & 1.27$^{+0.13}_{-0.10}$ &  & 548.62/592 \\
7 & IC~4329A & $<42.82$ & 1.69$\pm0.03$ & 153.94$^{+34.37}_{-24.85}$ & 0.12$^{+0.04}_{-0.05}$ & 2.48$^{+0.88}_{-0.59}$ & 2.68$^{+0.11}_{-0.29}$ & 2.80$^{+0.50}_{-0.75}$ & 6.11$^{+0.28}_{-0.26}$ & 4.58$^{+1.46}_{-0.96}$ & 1442.41/1335 \\
8 & MCG-6-30-15 &  & 2.06$\pm0.02$ & 149.73$^{+31.15}_{-12.46}$ & 1.04$^{+0.12}_{-0.18}$ & 2.98$^{+0.40}_{-0.33}$ & 1.69$^{+0.07}_{-0.22}$ & 2.51$^{+0.34}_{-0.36}$ & 2.46$^{+0.07}_{-0.03}$ & 1.31$^{+0.52}_{-0.87}$ & 933.30/878 \\
9 & Mkn~335 &  & 2.15$^{+0.10}_{-0.12}$ & $>194.58$ & 3.75$^{+0.83}_{-1.21}$ & 5.32$^{+2.05}_{-0.98}$ & $<2.35$ & 2.36$^{+0.82}_{-0.41}$ & 0.29$\pm0.02$ & 0.57$^{+0.48}_{-0.51}$ & 267.92/267 \\
10 & Mrk1040 &  & 1.83$^{+0.03}_{-0.04}$ & $>359.01$ & 0.52$^{+0.13}_{-0.11}$ & 2.23$^{+0.49}_{-0.64}$ & 2.76$^{+0.13}_{-0.45}$ & 2.71$^{+1.21}_{-1.11}$ & 1.75$^{+0.13}_{-0.23}$ &  & 581.81/578 \\
11 & Mrk~110 & 62.48$^{+43.30}_{-28.46}$ & 1.83$^{+0.07}_{-0.05}$ & 219.30$^{+437.39}_{-112.36}$ & 0.16$^{+0.10}_{-0.07}$ & $<2.35$ & 2.70$^{+0.06}_{-0.93}$ & $<6.54$ & 2.30$^{+0.39}_{-0.18}$ &  & 510.46/487 \\
12 & Mrk~590 &  & 1.64$^{+0.16}_{-0.12}$ & $>54.68$ & 0.51$^{+0.21}_{-0.20}$ & $<2.38$ & $<2.95$ & $>2.17$ & 0.19$^{+0.12}_{-0.04}$ &  & 191.93/198 \\
13 & Mrk~766 &  & 2.24$^{+0.06}_{-0.02}$ & $>455.64$ & 1.06$\pm0.19$ & 1.87$^{+0.48}_{-0.69}$ & $<2.42$ & 0.90$^{+0.16}_{-0.14}$ & 1.47$\pm0.07$ &  & 558.97/527 \\
14 & Swift~J2127.4+5654 &  & 2.13$^{+0.07}_{-0.08}$ & 138.0$^{+94.52}_{-39.64}$ & 1.78$^{+0.81}_{-0.44}$ & 1.86$^{+0.45}_{-0.43}$ & 1.77$^{+0.24}_{-0.38}$ & $<0.67$ & 1.58$^{+0.08}_{-0.07}$ &  & 664.01/633 \\
\hline
\hline
\end{tabular}
\end{center}
\end{table*}

\begin{table*}
\scriptsize
\tabcolsep 0.1cm
\caption{Best-fit parameters of the models {\sc constant$\times$tbabs$\times$relxill} and {\sc constant$\times$tbabs(relxill+xillver)} for \textit{i} = 30$^\circ$ \& $a=0.998$. $\Gamma$: hard X-ray photon index, $\beta$1: emissivity index, log~$\xi$: logarithm of ionisation parameter, $A_{\rm Fe}$: iron abundance in solar units, $E_{\rm cut}$: cut-off energy in keV, $R_f$: reflection fraction, $N_{\rm relxill}$: Normalisation of {\sc relxill}, $N_{\rm xillver}$: Normalisation of {\sc xillver}.}
\begin{center}
\label{tab_apt998}
\begin{tabular}{clcccccccccccccccccc}
\hline
\hline
\multicolumn{1}{c}{No.} & \multicolumn{1}{l}{Object} &  \multicolumn{1}{c}{$N_{\rm H}^{\rm Int}$} & \multicolumn{1}{c}{$\Gamma$} & \multicolumn{1}{c}{$E_{\rm cut}$} & \multicolumn{1}{c}{$R_f$} & \multicolumn{1}{c}{$\beta$1} & \multicolumn{1}{c}{log~$\xi$} & \multicolumn{1}{c}{$A_{\rm Fe}$} & \multicolumn{1}{c}{$N_{\rm relxill}$} & \multicolumn{1}{c}{$N_{\rm xillver}$} & \multicolumn{1}{c}{$\chi^2$/dof} \\
\\
\multicolumn{1}{c}{} & \multicolumn{1}{l}{} &  \multicolumn{1}{c}{10$^{20}{\rm cm}^{-2}$} &  \multicolumn{1}{c}{} & \multicolumn{1}{c}{keV} & \multicolumn{1}{c}{} & \multicolumn{1}{c}{} & \multicolumn{1}{c}{} & \multicolumn{1}{c}{} & \multicolumn{1}{c}{10$^{-4}$} & \multicolumn{1}{c}{10$^{-5}$} & \multicolumn{1}{c}{}
\\
\hline
1 & 3C~382 & 36.62$^{+54.12}_{-35.78}$ & 1.67$^{+0.05}_{-0.06}$ & 132.78$^{+99.90}_{-38.44}$ & 0.13$^{+0.07}_{-0.03}$ & $<2.00$ & $<2.14$ & $>4.39$ & 1.71$^{+0.28}_{-0.17}$ &   & 650.13/666\\
2 & 4C~74.26 &  & 1.80$\pm0.05$ & 154.11$^{+55.68}_{-44.28}$ & 0.61$\pm0.16$ & 2.32$^{+0.32}_{-0.37}$ & 2.69$^{+0.06}_{-0.35}$ & 1.39$^{+0.84}_{-0.67}$ & 1.35$\pm0.08$ &   & 696.51/664 \\
3 & Ark~120 &  & 1.97$\pm0.03$ & 393.61$^{+493.76}_{-153.24}$ & 0.62$^{+0.10}_{-0.12}$ & $<1.52$ & $<1.07$ & 2.44$^{+0.65}_{-0.57}$ & 2.36$^{+0.21}_{-0.16}$ &   & 693.28/647 \\
4 & Ark~564 & & 2.31$^{+0.05}_{-0.06}$ & 59.00$^{+14.12}_{-11.36}$ & 0.59$^{+0.40}_{-0.23}$ & $<1.91$ & $>4.27$ & 3.57$^{+3.64}_{-1.91}$ & 0.73$^{+0.31}_{-0.24}$ &   & 644.43/652 \\
5 & ESO~141-G~055 & & 1.92$\pm0.03$ & $>446.04$ & 0.52$^{+0.06}_{-0.08}$ & $<1.33$ & $<2.02$ & 1.58$^{+0.64}_{-0.55}$ & 1.72$^{+0.06}_{-0.08}$ &   & 669.78/635 \\
6 & Fairall~9 & & 1.92$\pm0.04$ & 336.75$^{+376.61}_{-128.54}$ & 0.66$^{+0.13}_{-0.14}$ & 1.44$^{+0.49}_{-1.20}$ & $<1.08$ & 3.03$^{+0.76}_{-0.67}$ & 1.27$^{+0.13}_{-0.10}$ &   & 548.35/592 \\
7 & IC~4329A & $<40.77$ & 1.70$^{+0.03}_{-0.04}$ & 156.89$^{+30.64}_{-27.66}$ & 0.13$\pm0.05$ & 2.32$^{+0.64}_{-0.44}$ & 2.69$^{+0.20}_{-0.30}$ & 2.83$^{+1.02}_{-0.69}$ & 6.11$^{+0.25}_{-0.29}$ & 4.64$^{+1.30}_{-1.10}$ & 1441.44/1335 \\
8 & MCG-6-30-15 & & 2.07$\pm0.03$ & 160.02$^{+28.76}_{-18.26}$ & 1.25$^{+0.17}_{-0.20}$ & 2.53$^{+0.26}_{-0.10}$ & 1.69$^{+0.07}_{-0.17}$ & 2.94$^{+0.54}_{-0.63}$ & 2.45$\pm0.07$ & $<1.33$ & 938.16/878 \\
9 & Mkn~335 & & 1.87$^{+0.26}_{-0.10}$ & 128.94$^{+732.96}_{-57.63}$ & 4.85$^{+1.45}_{-1.14}$ & 3.97$^{+0.40}_{-0.47}$ & 2.32$^{+0.56}_{-1.03}$ & $>3.76$ & 0.22$^{+0.03}_{-0.04}$ & 0.35$^{+0.42}_{-0.16}$ & 262.83/267 \\
10 & Mrk~1040 &  & 1.83$^{+0.02}_{-0.04}$ & $>355.73$ & 0.56$^{+0.15}_{-0.12}$ & 2.11$^{+0.35}_{-0.55}$ & 2.77$^{+0.15}_{-0.34}$ & 2.91$^{+1.39}_{-1.22}$ & 1.73$^{+0.21}_{-0.23}$ &   & 581.85/578 \\
11 & Mrk~110 & 62.43$^{+43.99}_{-19.96}$ & 1.83$^{+0.07}_{-0.08}$ & 219.30$^{+412.27}_{-113.95}$ & 0.16$^{+0.11}_{-0.07}$ & $<2.26$ & $<2.75$ & $<3.57$ & 2.30$^{+0.39}_{-0.18}$ &   & 510.48/487 \\
12 & Mrk~590 &   & 1.62$^{+0.19}_{-0.12}$ & $>53.41$ & 0.49$^{+0.29}_{-0.18}$ & $<2.18$ & $<2.66$ & $>1.39$ & 0.18$^{+0.13}_{-0.04}$ &   & 192.10/198 \\
13 & Mrk~766 &   & 2.28$^{+0.07}_{-0.06}$ & $>354.63$ & 1.26$^{+0.20}_{-0.23}$ & $<1.91$ & 1.69$^{+0.08}_{-0.17}$ & 0.71$^{+0.16}_{-0.14}$ & 1.47$^{+0.08}_{-0.09}$ &   & 549.90/527 \\
14 & Swift~J2127.4+5654 & & 2.14$^{+0.08}_{-0.10}$ & 126.75$^{+36.55}_{-34.99}$ & 1.86$^{+1.10}_{-0.48}$ & 1.66$^{+0.41}_{-0.77}$ & 1.58$^{+0.44}_{-0.22}$ & $<0.72$ & 1.60$\pm0.12$ &   & 664.73/633 \\
\hline
\hline
\end{tabular}
\end{center}
\end{table*}

\section{Results \& Discussion}
\label{sec4}

The present work determines the relativistic reflection fraction of a sample of 14 type~1 Seyferts observed with \textit{NuSTAR}. In this sample, the spectra of 11 sources were modelled with {\sc relxill}, and those of three sources with a combination of {\sc relxill} and {\sc xillver}. We fitted the spectra with the same model for spin parameters of 0 and 0.998. For both the spin parameters, the photon index and reflection fraction were well constrained for all the sources. The values of $R_{f}$ obtained for these sources are roughly in the range of $0.1-3.8$, with a mean value of around 0.9 for $a=0$. For $a=0.998$, $R_{f}$ ranges from $\sim$0.1 to $\sim$4.9 and has a mean value of $\sim$0.98. However, we find that both the reflection fraction and the photon index are consistent within errorbars for non-rotating ($a=0$) and maximally rotating ($a=0.998$) black holes. The obtained reflection fraction is similar to the previous results from \cite{2011ApJ...728...58B} and \cite{2014MNRAS.437.2845B} and higher than the values reported by \cite{2017ApJ...849...57D} and \cite{2011A&A...532A.102R}. However, the inconsistency in the values of reflection fraction may be attributed to the different models used in the analysis. The above mentioned studies used {\sc pexrav} and {\sc pexmon} models for describing the reflection component while the present work used {\sc relxill} model.

A summary of the broadband spectral analysis of these sources are given in Table~\ref{tab_a0} and Table~\ref{tab_apt998}. The broadband spectral fitting plot for IC~4329A using the model {\sc constant$\times$tbabs$\times$ztbabs(relxill + xillver)} is shown in Fig.~\ref{Fig_fit}. The spectral fit and residuals for the final model for the whole sample are given in Appendix~\ref{B}.

\subsection{Correlations}

\begin{figure*}
\begin{center}
\includegraphics[trim=0cm 0cm 0cm 0cm, clip=true, width =8.81cm, angle=0]{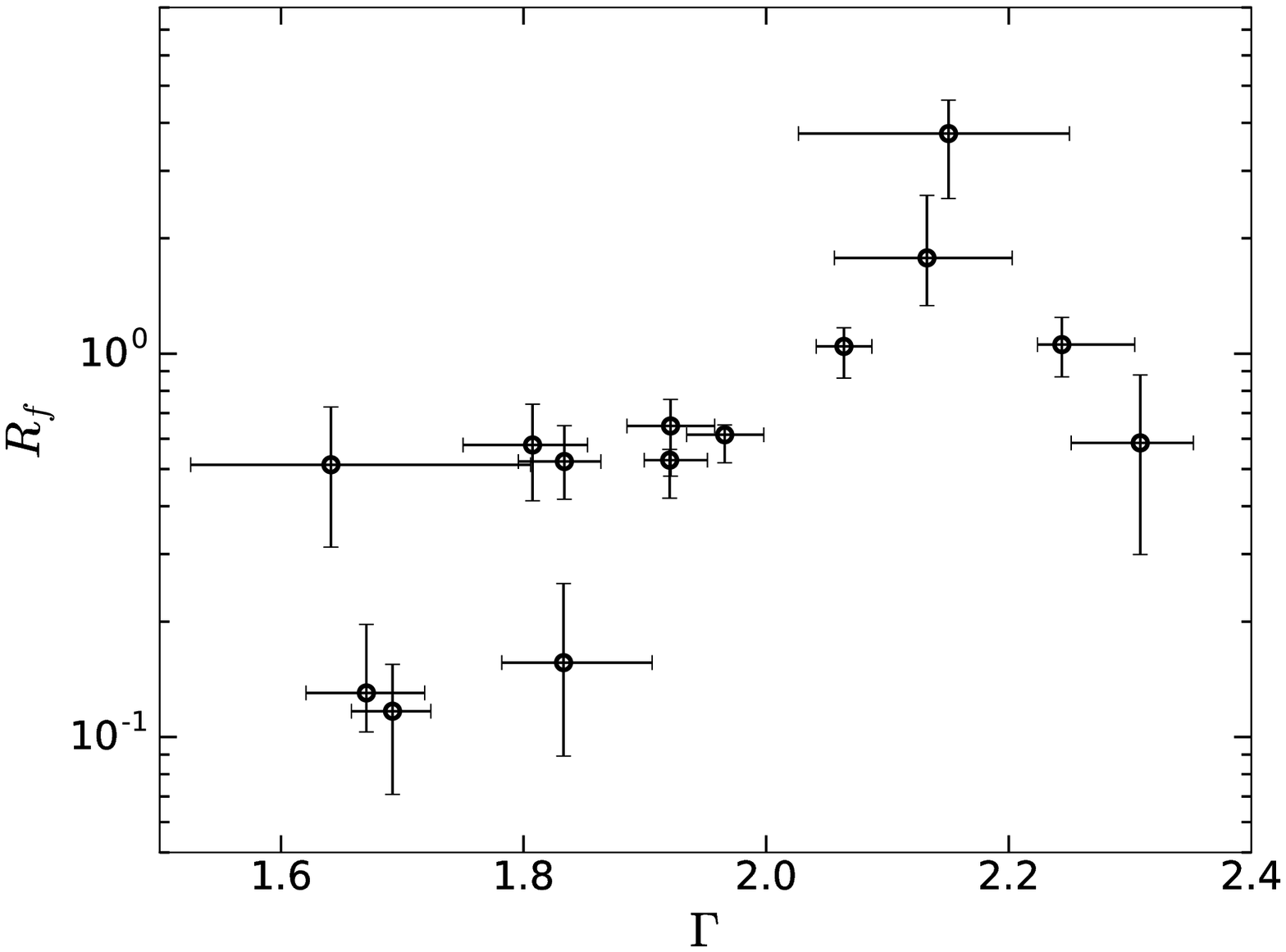}
\includegraphics[trim=0cm 0cm 0cm 0cm, clip=true, width =8.81cm, angle=0]{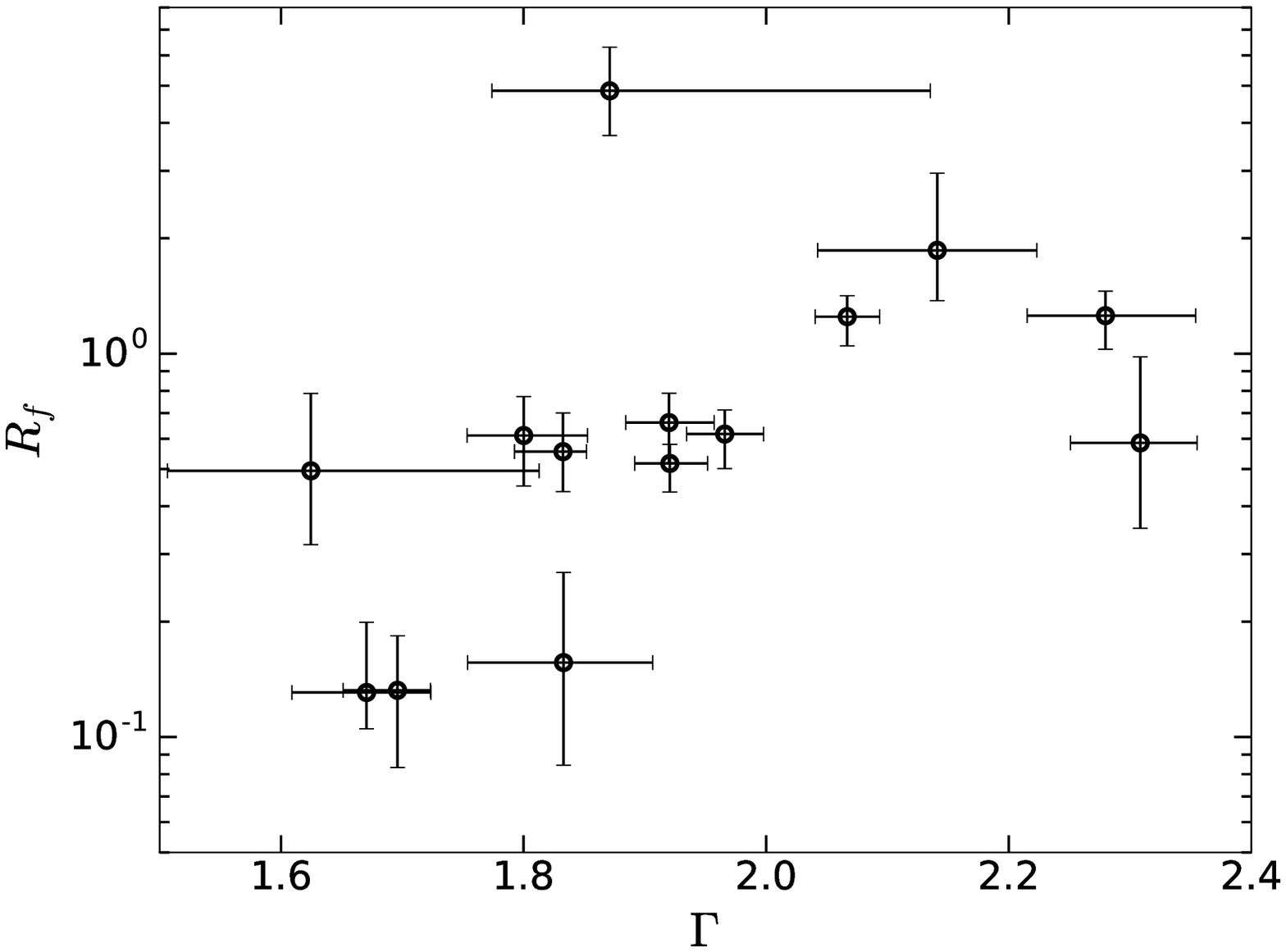}
\caption[]{Relationship between $R_f$ and $\Gamma$ for the sample for $a$=0 (left) and $a$=0.998 (right).}
\label{Fig_Rgamma}
\end{center}
\end{figure*}

\begin{figure*}
\begin{center}
\includegraphics[trim=0cm 0cm 0cm 0cm, clip=true, width =8.7cm, angle=0]{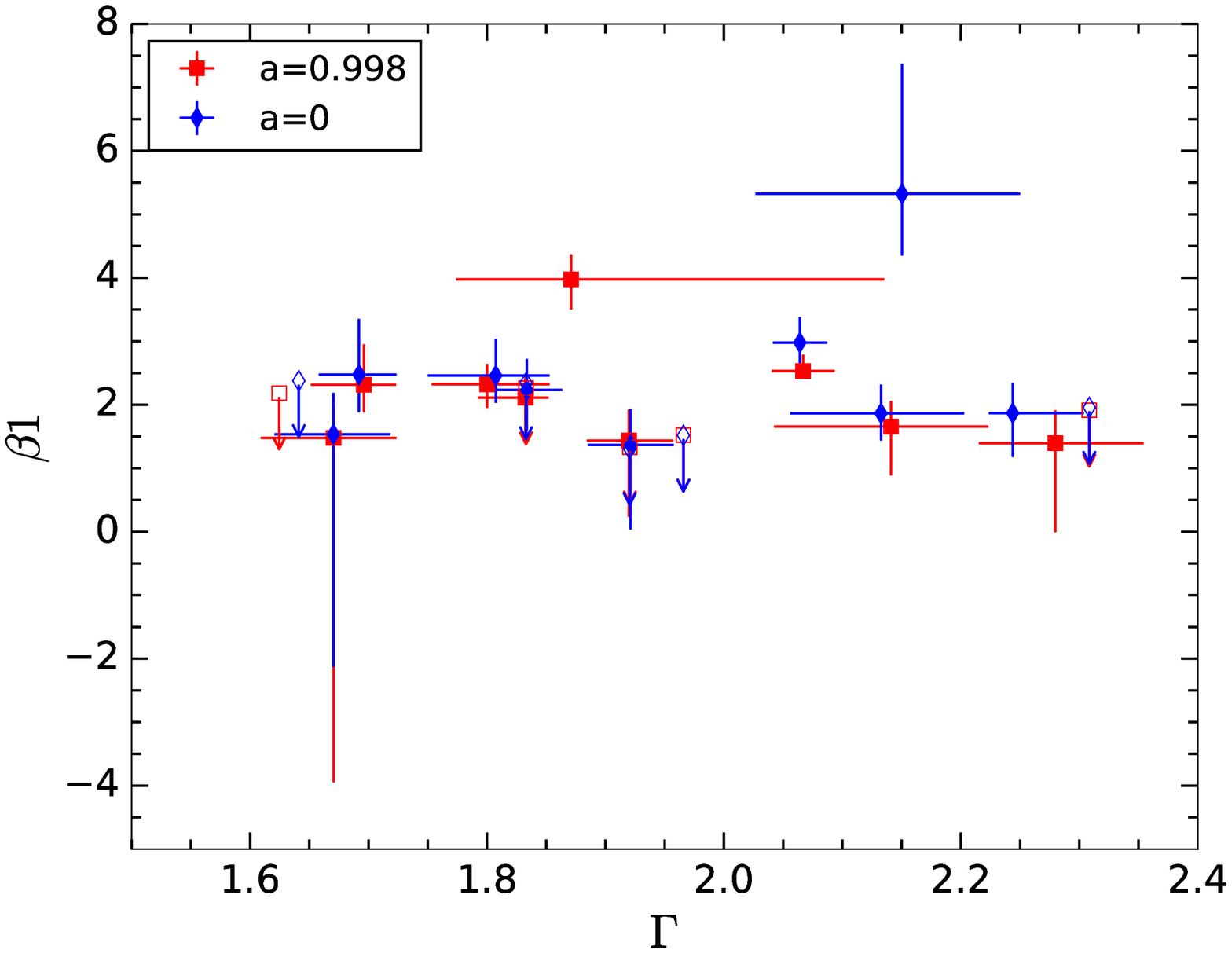}
\includegraphics[trim=0cm 0cm 0cm 0cm, clip=true, width =8.5cm, angle=0]{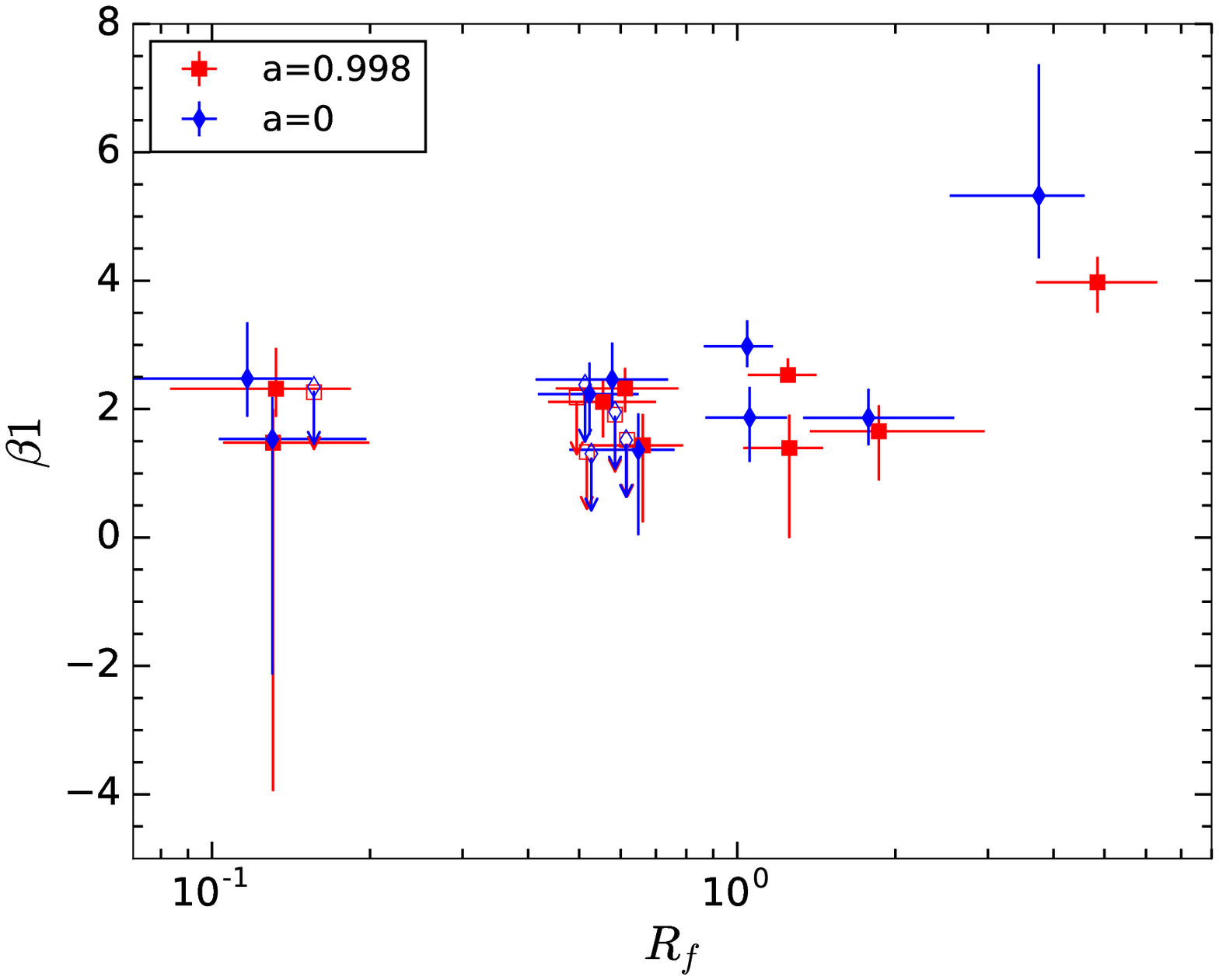}
\caption[]{Plot showing emissivity index $\beta$1 vs. $\Gamma$ (left) and $\beta$1 vs. $R_f$ (right) for the sample. Filled red squares represent the values for $a$=0.998, and blue filled diamond-shaped data points represent the same for $a$=0. In case $\beta$1 is not constrained in the fit, the upper limit for the parameter values are denoted by open squares/diamonds with lower arrows.}
\label{Fig_index}
\end{center}
\end{figure*}

We obtained the correlations between the relativistic reflection fraction and the photon index in our sample using Spearman's rank-order method \cite{numerical_recipes}. A significant positive correlation is observed between $R_{f}$ and $\Gamma$ with a ${\rm rank}$ of 0.83 and p-value of 0.0003 for $a=0$. A similar trend is observed for $a=0.998$ as well with a ${\rm rank}$ of 0.65 and a p-value of 0.01. The parameters for different spin values are plotted in Fig.~\ref{Fig_Rgamma}. Pearson's correlation also shows a weak positive linear trend between log$R_{f}$ and $\Gamma$ with a null hypothesis probability (p-value) of 0.05 and a rank of 0.54 for a=0. We have also checked the correlations between other parameters such as X-ray luminosity, X-ray Eddington ratio, Fe abundance, ionisation parameter and high energy cut-off. However, we did not find any significant correlations among these parameters.

Many previous works have studied the dependence of Compton reflection on the hard X-ray spectral slope in AGN (e.g. \citet{1999MNRAS.303L..11Z,1999ApJ...510L.123B,2001MNRAS.326..417M,2007ApJ...664..101M,2008A&A...485..417D,
2009MNRAS.399.1293M,2016A&A...588A..70B,2017ApJ...849...57D,2018ApJ...854...33Z}). \citet{1999MNRAS.303L..11Z} found a significant correlation between the relative strength of Compton reflection (defined as $\Omega/2\pi$; $\Omega$ is the solid angle subtended by the reflector) and $\Gamma$, in their sample of Seyferts. They argued that an internal feedback mechanism, where the medium emitting seed photons for the primary X-ray emission also serves as the medium for reflection, is responsible for the observed correlation. \citet{2007ApJ...664..101M}, though observed a strong correlation between the relative amount of reflection and photon index in a sample of \textit{RXTE}-observed Seyfert 1 and 1.2 galaxies, ruled out any physical relevance arguing that it is due to the presence of model degeneracies. \citet{2008A&A...485..417D} found a significant correlation between the relative amount of reflection and the photon index in a sample of Seyfert galaxies in the local Universe ($z\leq0.1$). Contrary to this, \citet{2009MNRAS.399.1293M} did not observe any correlation between reflection fraction and $\Gamma$ in a sample of type~1 AGN observed with \textit{INTEGRAL}. Another study by \citet{2016MNRAS.458.2454L} on the hard X-ray spectra of an \textit{INTEGRAL} sample of 28 Seyfert galaxies, together with the X-ray data from \textit{XMM-Newton}, \textit{Suzaku} and \textit{RXTE}, reported a less prominent correlation between the reflection strength and the photon index. Here, using the better quality data from \textit{NuSTAR}, we further confirm the $R_{f}-\Gamma$ correlation in our sample of Seyfert~1s. Recently, \cite{2019A&A...626A..40P} studied the \textit{NuSTAR} sample of local AGN by classifying the sources based on X-ray spectral shape. They obtained an average reflection strength (measure of Compton hump intensity with respect to primary emission) of about 1.7 and 0.5, respectively for mildly obscured (23 $<$ log$N_{\rm H} <$ 24) and lightly obscured (21 $<$ log$N_{\rm H} <$ 22) sources. For unabsorbed sources, they observed that the reflection strength is correlated to the photon index.

The correlation between relativistic reflection fraction and photon index is significant in our sample, and this can be explained as follows. The primary X-ray emission is produced by the inverse Compton scattering
of the accretion disc photons by the hot electrons in the corona, and a portion of the same X-ray photons irradiates the disc producing the reflection features. Also, the slope of the power law is directly linked to the rate of cooling of the hot corona. Considering such a scenario, the more the input seed photons entering the corona, the stronger the cooling of plasma which results in a steeper X-ray power law. This results in a higher fraction of X-ray photons illuminating the accretion disc leading to larger reflection fraction. Since $R_{f}$ represents the intensity of the primary X-ray source irradiating the inner regions of the accretion disc relative to the intensity directly reaching the observer, the parameter, in turn, depends on the relative geometry of the accretion disc and the corona. Thus the observed $R_{f}-\Gamma$ correlation could be a consequence of the change in the disc-corona geometry.

We also fitted the spectra of the sources IC~4329A and Swift~J2127.4+5654 by varying the inner disc radius instead of fixing it at the innermost stable orbit. However, $R_{\rm in}$ was not constrained, and we do not find any change in the reflection fraction. 

Here, we are using the standard version of the {\sc relxill} model that assumes the illumination profile of the accretion disc as a broken power law. In this model, the geometry of the illuminating source is not well defined. However, the emissivity index of the profile is expected to vary with different source heights and accretion disc radii. Hence we can elucidate the geometry of the system from the emissivity index. Nevertheless, the emissivity index for the sample does not vary and is not constrained for a few sources. The plots for emissivity index versus $\Gamma$ and $R_f$ are shown in Fig.~\ref{Fig_index}. However, the lamp post flavour of the model {\sc relxilllp} has a well-specified geometry where an on-axis primary source is located above the black hole at a height \textit{h}. We explored the possibility of constraining the geometry of the system using this model. We fitted the spectra of the sources IC~4329A and Swift~J2127.4+5654 with {\sc relxilllp} and obtained the source height in units of gravitational radius $R_{\rm g}$. The height of the corona in Swift~J2127.4+5654 is found to be $\sim$39(50)~$R_{\rm g}$ for $a=0$(0.998) and in the case of IC~4329A the parameter is poorly constrained and has an upper limit of 23(22)~$R_{\rm g}$. The inner disc radius $R_{\rm in}$ in both the sources are not constrained. We also note that neither the reflection fraction nor the photon index has changed significantly from the values we obtained for {\sc relxill} model. 

For assessing the exact geometry of the system, the current data and model seem to be inadequate. More appropriate modelling of AGN which can properly constrain the accretion disc and coronal geometry can reveal how the reflection fraction from the inner disc is determined by the geometry of the sources.

\section{Conclusion}
\label{sec5}

In this work, we analysed the reflection spectra of a sample of 14 Seyfert type~1 galaxies using the X-ray data from the \textit{NuSTAR} observations. The X-ray spectra in the 3$-$79~keV band were modelled with {\sc relxill} that explains the relativistic reflection features along with the primary X-ray emission, and the {\sc xillver} component was used to model the features due to distant reflection. The relativistic reflection fraction $R_{f}$ of the sample was obtained from the model {\sc relxill} and is found to range from $\sim$0.12(0.13) to $\sim$3.75(4.85) for $a=0(0.998)$. The parameters $R_{f}$ and $\Gamma$ show a significant positive correlation with a rank of 0.83(0.65) (p-value~$\sim$~0.0003(0.01)). Since the slope of the X-ray power law is related to the rate of cooling of the plasma, steeper X-ray spectra indicate stronger cooling by the seed photons. The larger the area covered by the accretion disc as seen from the corona, the higher will be the seed photons entering the corona. This results in the steepening of the X-ray spectrum. Since the same accretion disc is responsible for the reflected emission, the larger area covered by the medium consequently enhances the reflection fraction. The observed $R_{f}-\Gamma$ correlation is thus resulting from the change in disc-corona geometry of the AGN in our sample.

\section{Acknowledgement}

We thank the anonymous referee for the useful comments and suggestions. This research has made use of data and/or software provided by the High Energy Astrophysics Science Archive Research Center (HEASARC), which is a service of the Astrophysics Science Division at NASA/GSFC and the High Energy Astrophysics Division of the Smithsonian Astrophysical Observatory. This work has made use of data obtained from the \textit{NuSTAR} mission, a project led by Caltech, funded by NASA and managed by NASA/JPL, and has utilised the {\sc nustardas} software package, jointly developed by the ASDC, Italy and Caltech, USA. This research has made use of the NASA/IPAC Extragalactic Database (NED) which is operated by the Jet Propulsion Laboratory, California Institute of Technology, under contract with the National Aeronautics and Space Administration. We acknowledge the usage of the HyperLeda database ({\url{http://leda.univ-lyon1.fr}). This research has made use of the SIMBAD database, operated at CDS, Strasbourg, France.

\def\aj{AJ}%
\def\actaa{Acta Astron.}%
\def\araa{ARA\&A}%
\def\apj{ApJ}%
\def\apjl{ApJ}%
\def\apjs{ApJS}%
\def\ao{Appl.~Opt.}%
\def\apss{Ap\&SS}%
\def\aap{A\&A}%
\def\aapr{A\&A~Rev.}%
\def\aaps{A\&AS}%
\def\azh{AZh}%
\def\baas{BAAS}%
\def\bac{Bull. astr. Inst. Czechosl.}%
\def\caa{Chinese Astron. Astrophys.}%
\def\cjaa{Chinese J. Astron. Astrophys.}%
\def\icarus{Icarus}%
\def\jcap{J. Cosmology Astropart. Phys.}%
\def\jrasc{JRASC}%
\def\mnras{MNRAS}%
\def\memras{MmRAS}%
\def\na{New A}%
\def\nar{New A Rev.}%
\def\pasa{PASA}%
\def\pra{Phys.~Rev.~A}%
\def\prb{Phys.~Rev.~B}%
\def\prc{Phys.~Rev.~C}%
\def\prd{Phys.~Rev.~D}%
\def\pre{Phys.~Rev.~E}%
\def\prl{Phys.~Rev.~Lett.}%
\def\pasp{PASP}%
\def\pasj{PASJ}%
\def\qjras{QJRAS}
\def\rmxaa{Rev. Mexicana Astron. Astrofis.}%
\def\skytel{S\&T}%
\def\solphys{Sol.~Phys.}%
\def\sovast{Soviet~Ast.}%
\def\ssr{Space~Sci.~Rev.}%
\def\zap{ZAp}%
\def\nat{Nature}%
\def\iaucirc{IAU~Circ.}%
\def\aplett{Astrophys.~Lett.}%
\def\apspr{Astrophys.~Space~Phys.~Res.}%
\def\bain{Bull.~Astron.~Inst.~Netherlands}%
\def\fcp{Fund.~Cosmic~Phys.}%
\def\gca{Geochim.~Cosmochim.~Acta}%
\def\grl{Geophys.~Res.~Lett.}%
\def\jcp{J.~Chem.~Phys.}%
\def\jgr{J.~Geophys.~Res.}%
\def\jqsrt{J.~Quant.~Spec.~Radiat.~Transf.}%
\def\memsai{Mem.~Soc.~Astron.~Italiana}%
\def\nphysa{Nucl.~Phys.~A}%
\def\physrep{Phys.~Rep.}%
\def\physscr{Phys.~Scr}%
\def\planss{Planet.~Space~Sci.}%
\def\procspie{Proc.~SPIED}%
\let\astap=\aap
\let\apjlett=\apjl
\let\apjsupp=\apjs
\let\applopt=\ao
\bibliographystyle{mnras} 
\bibliography{ref.bib} 



\appendix

\section{Residual plots}
\label{A}

The residuals for the spectral fits using the model {\sc constant$\times$tbabs$\times$powerlaw} is shown in Fig.~\ref{A_res1}.

\begin{figure*}
\begin{center}
\includegraphics[trim=0cm 0cm 0cm 0cm, clip=true, scale=1.0, angle=0]{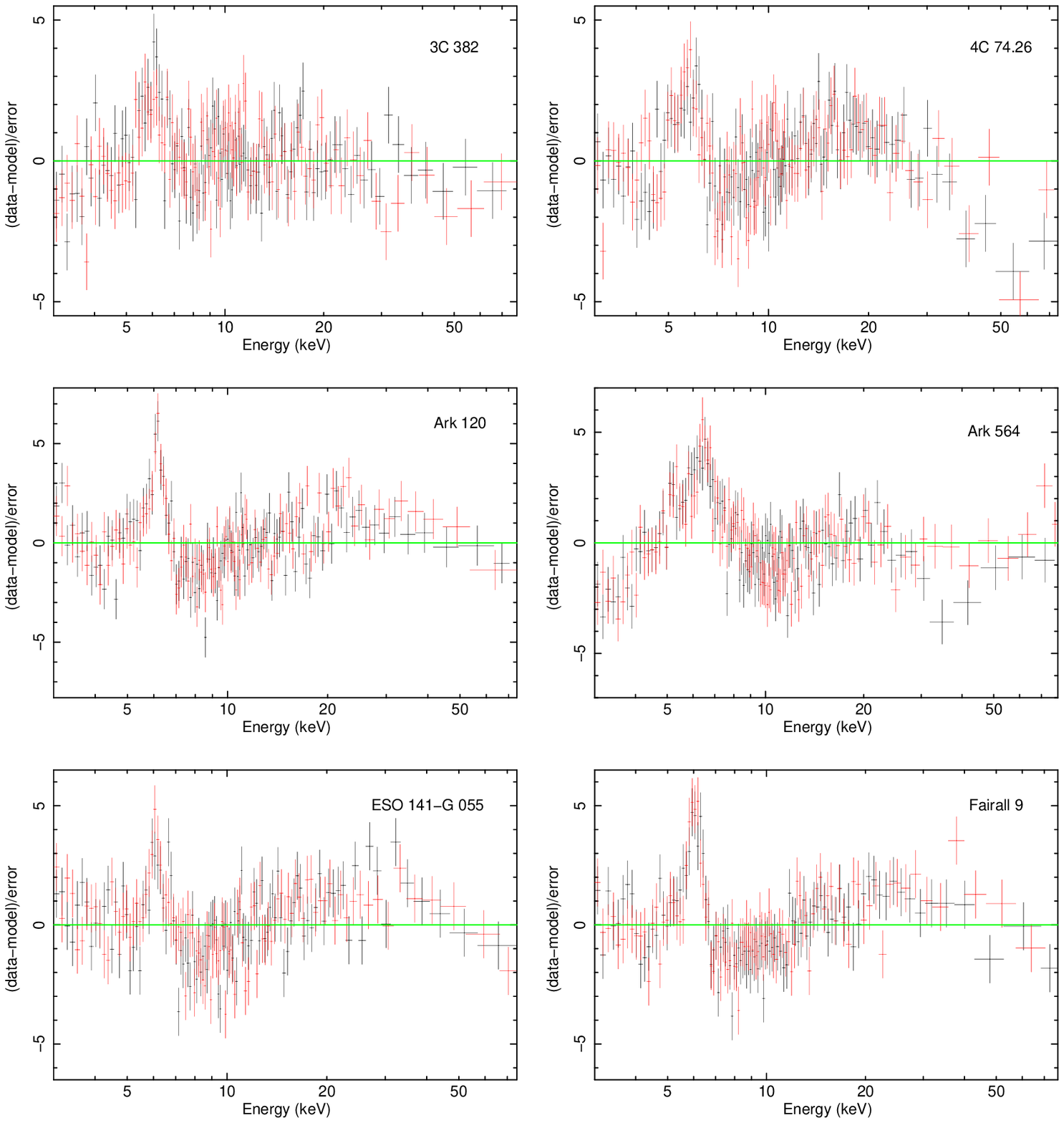}
\caption{Plots showing the residuals for the model {\sc constant$\times$tbabs$\times$powerlaw} fitted over the 3--79~keV range. All the data have been rebinned for plotting purpose.}
\label{A_res1}
\end{center}
\end{figure*}

\begin{figure*}
\begin{center}
\includegraphics[scale=1.0, angle=0]{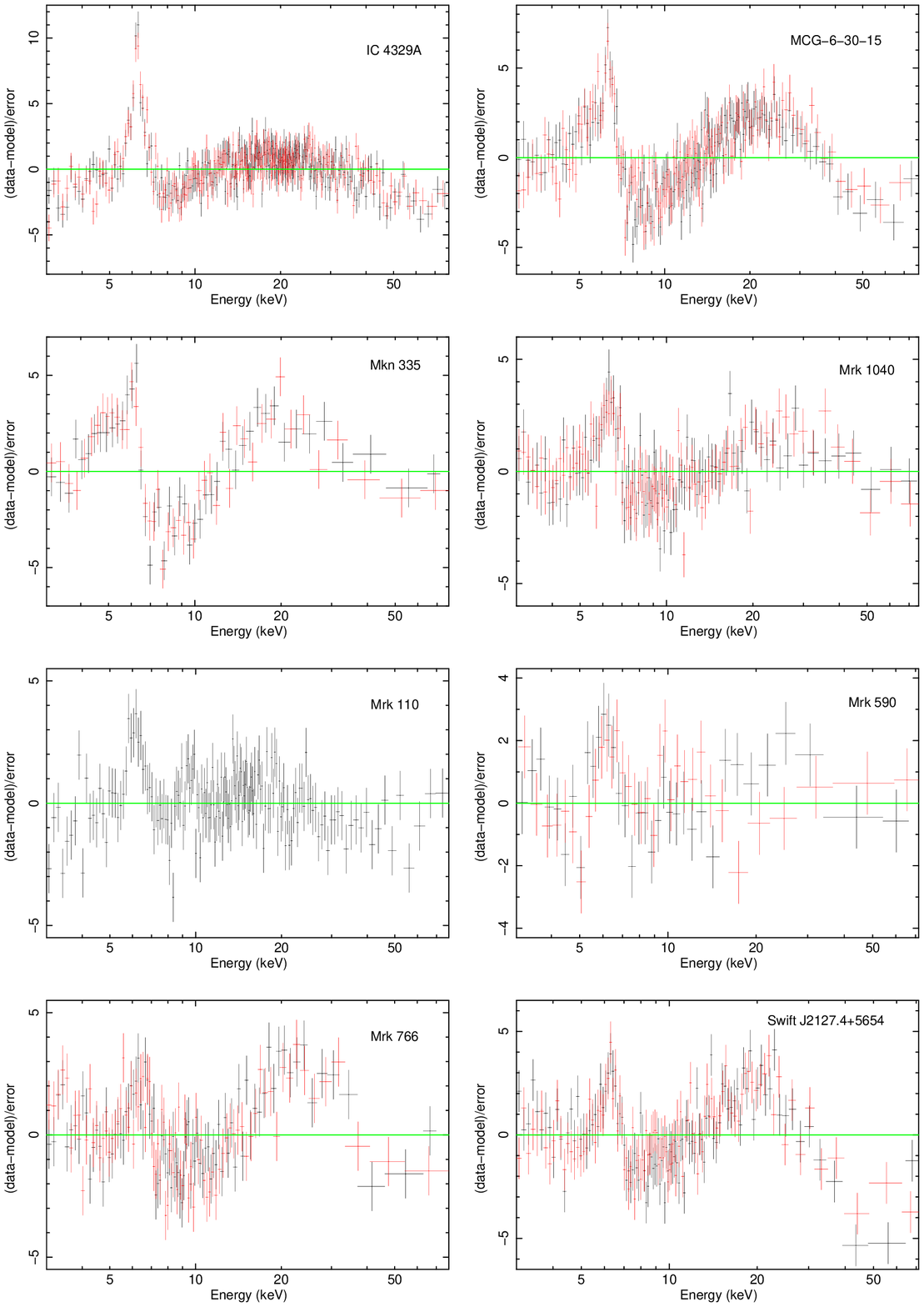}
\captionsetup{labelformat=empty}
\begin{center}
Figure~\ref{A_res1} (continued)
\end{center}
\label{A_res2}
\end{center}
\end{figure*}

\section{Spectral fitting plots}
\label{B}

The broadband spectral fits and residuals using the final best-fit model (for $a=0$) for the whole sample is plotted in Fig.~\ref{B_specfit1}.

\begin{figure*}
\begin{center}
\includegraphics[trim=0cm 0cm 0cm 0cm, clip=true, scale=0.85, angle=0]{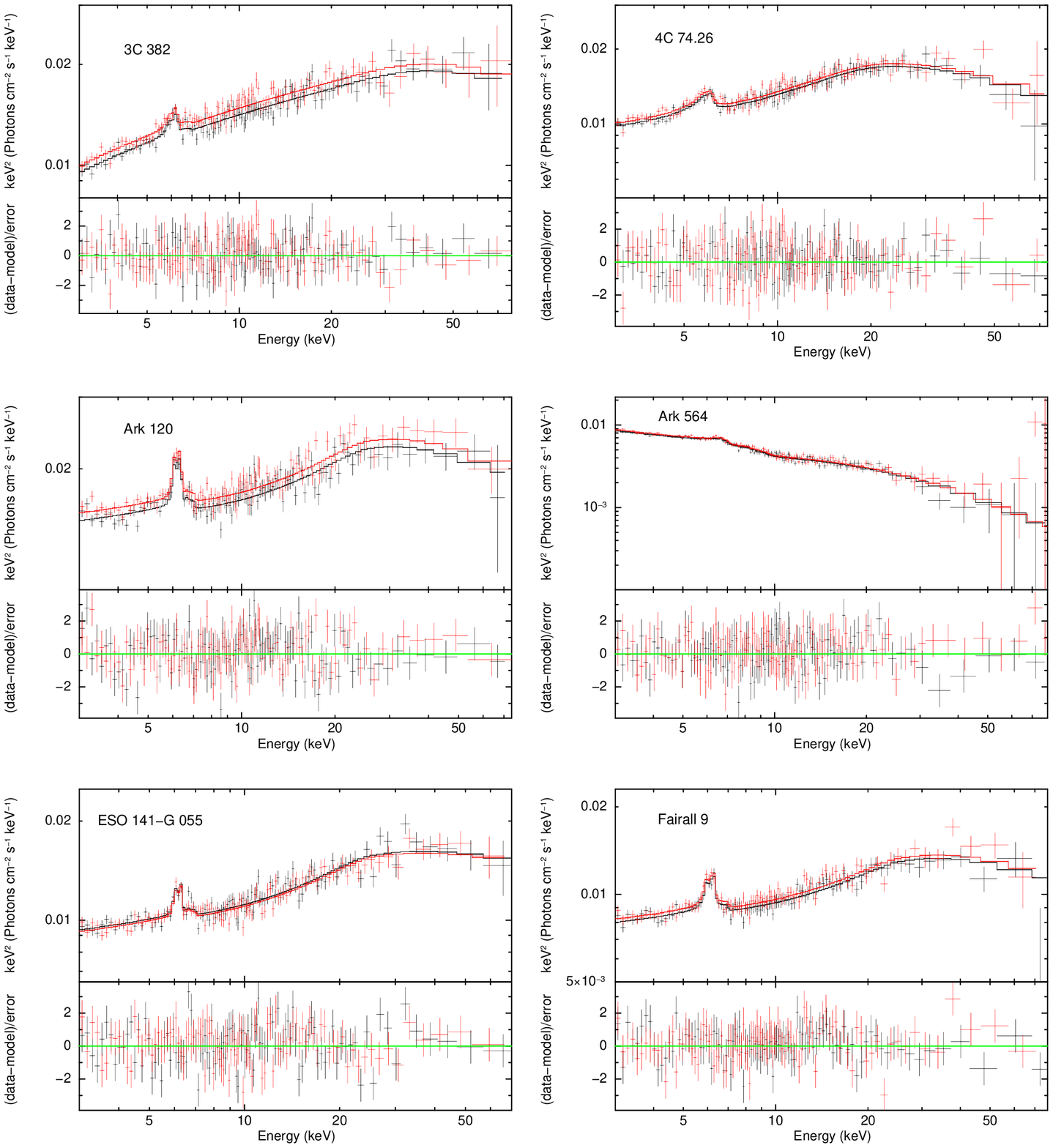}
\caption{Spectral fitting plots for the sample. Upper panels: The unfolded spectra and the best-fit models ({\sc constant$\times$tbabs$\times$relxill} or {\sc constant$\times$tbabs(relxill+xillver)}). Lower panels: The residuals of the spectral fit.}
\label{B_specfit1}
\end{center}
\end{figure*}

\begin{figure*}
\begin{center}
\includegraphics[scale=0.85, angle=0]{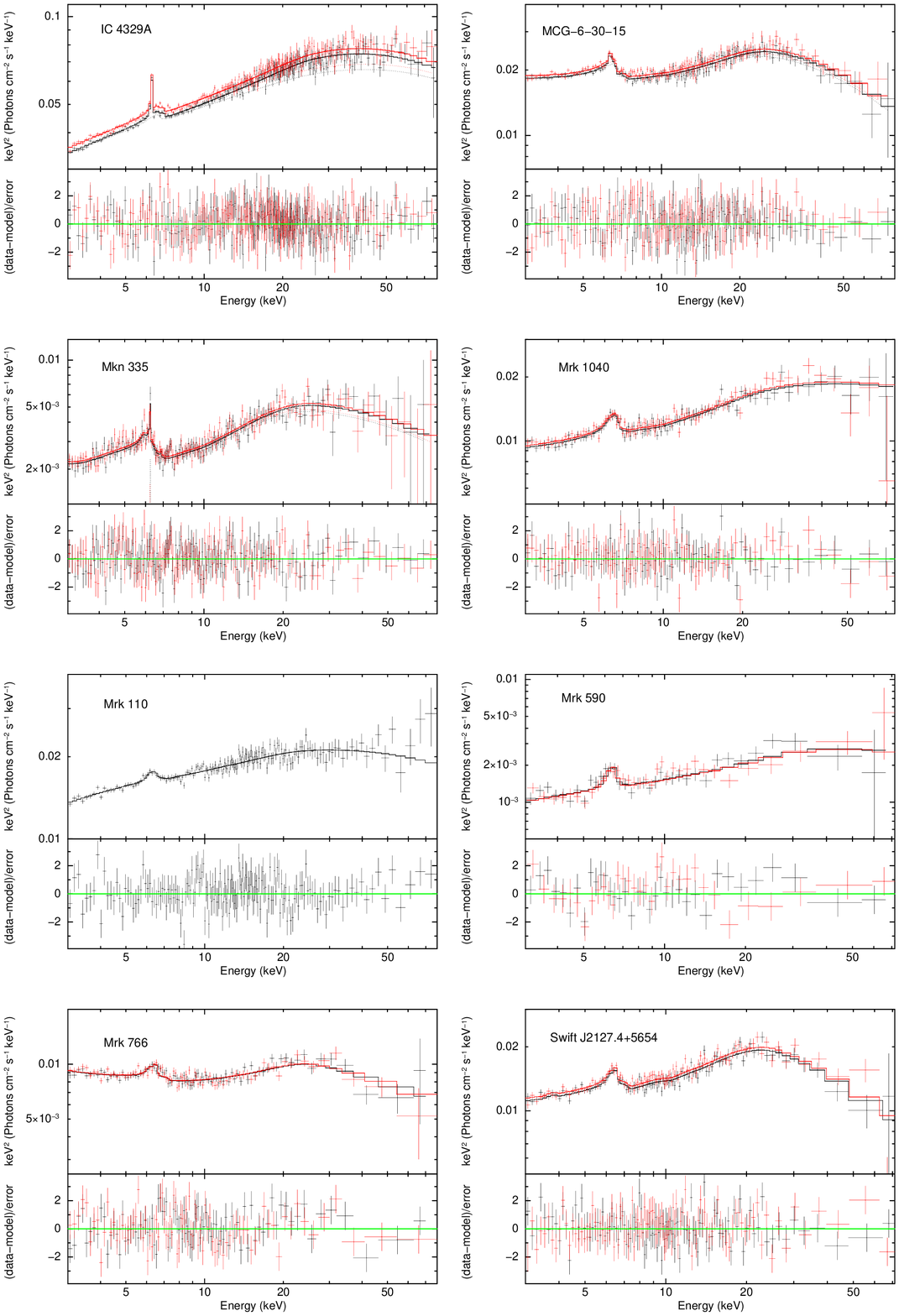}
\captionsetup{labelformat=empty}
\begin{center}
Figure~\ref{B_specfit1} (continued)
\end{center}
\label{B_specfit2}
\end{center}
\end{figure*}


\bsp	
\label{lastpage}
\end{document}